\documentclass[epj]{svjour}
\usepackage[utf8]{inputenc}
\usepackage[T1]{fontenc}
\date{\today}
\usepackage{amsfonts}
\usepackage{amsmath}
\usepackage{amssymb}
\usepackage{graphicx}
\usepackage{tabularx}%
\usepackage{overpic}
\usepackage{color}
\usepackage{mathrsfs}
\usepackage{subfigure}
\usepackage{color}

\newcommand{\dd}{\mathrm{d}}
\newcommand{\K}{\overline{K}}
\newcommand{\N}{\overline{N}}
\newcommand{\br}{\mathbf{r}}
\newcommand{\deron}[2]{\left(\frac{\partial{#1}}{\partial{#2}}\right)}
\newcommand{\deronb}[2]{\partial_{#2}{#1}}
\newcommand{\revision}[1]{}

\DeclareMathOperator{\e}{e}

\begin{document}\sloppy
\title{Statistical Mechanics of Two-dimensional Foams: %testing/probing/validity of the foundations/hypotheses OU testing the validity of the foundations.}
Physical Foundations of the Model.}

\author{Marc Durand}

\institute{Mati\`{e}re et Syst\`{e}mes Complexes (MSC), UMR 7057 CNRS \& Universit\'{e}
Paris Diderot, 10 rue Alice Domon et L\'{e}onie Duquet, 75205 Paris Cedex 13, France}

\abstract{
In a recent series of papers \cite{Durand1,Durand2,Durand3}, a statistical model that accounts for correlations between topological and geometrical properties of a two-dimensional shuffled foam has been proposed and compared with experimental and numerical data.
Here, the various assumptions on which the model is based are exposed and justified: the equiprobability hypothesis of the foam configurations is argued. The range of correlations between bubbles is discussed, and the mean field approximation that is used in the model is detailed. The two self-consistency equations associated with this mean field description can be interpreted as the conservation laws of number of sides and bubble curvature, respectively. Finally, the use of a ``Grand-Canonical'' description, in which the foam constitutes a reservoir of sides and curvature, is justified. %The limitations of the model are clearly stated.
  }

\maketitle

\section{Introduction}

Foams, granular materials, biological systems, or glasses share the common feature of being out-of-equilibrium systems from the point of view of standard statistical mechanics: either these systems possess a large number of metastable states in which they are trapped within a usual range of temperature (e.g.: foam or granular materials at rest, glassy systems); or a flux of energy is continuously injected from non-thermal sources and is lost through internal friction, allowing them to explore their
phase space (e.g.: sheared or tapped granular systems, biological cells).
As these systems involve many individual objects (bubbles, grains, cells, or atoms), there is
a strong motivation to treat them with thermodynamic methods.
The pioneering
work of Edwards on granular matter \cite{Edwards-and-co,Blumenfeld} has shown how the
powerful arsenal of statistical physics can be extended to
out-of-equilibrium systems. 
Because of the perfect space-filling property of a foam (\textit{i.e.} without gaps nor overlaps), and the high deformability of its constituents (bubbles), its metastable states are described by a set of well established rules \cite{WeaireBook1,livre_mousse}. For that reason, foams are very good candidates of out-of-equilibrium model systems.

\revision{
Various models using the concepts of statistical thermodynamics have been proposed to describe the geometrical and topological properties of 2D foams  \cite{Schliecker1,Schliecker2,Rivier1,Rivier2,Fortes3,Iglesias,sire}.
These models either use an ad-hoc interaction potential between bubbles \cite{Schliecker1,Schliecker2}, involve (rather than deduce) empirical laws correlating size and side distributions \cite{Rivier1,Rivier2,Fortes3}, or seek the distribution of bubble size that follows some optimization scheme, which is relevant at times long enough for the size distribution to vary (\textit{e.g.}, under coarsening or bubble coalescence) \cite{Rivier1,Rivier2,Iglesias,sire}. 
}
Recently, a new statistical description has been proposed to describe the structural properties of a two-dimensional (2D) dry foam \cite{Durand1}. The model succeeded to reproduce, without any adjusting parameter, the correlation between geometry (distribution of bubble size) and topology (distribution of side number) for foams with moderate size dispersities \revision{which are shuffled, either mechanically or thermally (using extended Potts simulations)} \cite{Durand2,Durand3}.
No semi-empirical laws (such as Aboav-Weaire or Lewis laws \cite{WeaireBook1,livre_mousse}) are used in this approach; instead, laws that correlate geometry and topology properties for individual bubbles and for the entire foam are deduced from the model. 
Interplay between geometrical and topological features of a foam is crucial in determining  \textit{e.g.} rheological properties or coarsening rate. Although the primary goal of this approach is to describe the statistical properties of foams which are uniformly shuffled, it may also describe disorders in foams at rest, if the foam production process involves uniform shuffling. Furthermore, because of the resemblance of many epithelial tissues with densely packed soap bubbles \cite{PNAS}, this approach could be extended to biological tissues.
Although first comparisons with numerical and experimental data show a remarkable agreement \cite{Durand2,Durand3}, this theoretical model is based on successive assumptions which need to be tested. 

The aim of the present paper is to establish solid foundations to this theoretical framework, by discussing and testing the validity of the different underlying hypotheses. In particular, the mean field approach used in the model is detailed, and one shows that the associated self-consistency equations can be interpreted as the conservation equations of two scalar quantities.
The validity of the hypotheses are tested with simulations of shuffled foams. Simulations are well suited because they allow to have precise measures on the pressure inside every bubble and curvature of their sides, which cannot be easily done in experiments.

The structure of the paper is as follows: Section \ref{section_predictions} summarizes the main results of the model and its success to reproduce experimental and numerical data.
Section \ref{section_constraints} reviews the geometrical, topological and mechanical laws that govern the structure of a foam at rest. Because of these space-filling constraints, bubble shapes are correlated.
In Section \ref{section_shuffling}, mechanical and thermal shufflings are defined and compared, and the postulate of equiprobability of the visited states is discussed.
Section \ref{section_MFA} presents a mean field theory in which correlations between neighbouring bubbles are disregarded, but correlation between the geometry (size) of a bubble and its topology (number of sides) is taken into account. Within the frame of this approximation, two \textit{invariants} are identified in a 2D shuffled foam, \textit{i.e.} quantities that are exchanged by the bubbles but whose sum is preserved through the elementary topological process of neighbour swapping (T1 event). The mean field approximation is then tested with numerical simulations.
Section \ref{section_stat_mech} discusses the validity of the assumptions made to use a grand-canonical description: the short correlation length between the bubbles, and the extensivity of the state variables. Finally, Section \ref{section_outlook} discusses possible extensions of the model to three-dimensional foams and other cellular systems.

\section{Main results of the model} \label{section_predictions}
In this section, one briefly recalls the main predictions of the model, and its success to reproduce, without any adjusting parameter, the correlations between topological and geometrical properties of a foam.
The model concerns soft (or liquid), two-dimensional foams in the low density limit, so bubbles have polygonal shapes; their structures are governed by space-filling constraints that are detailed in Sect. \ref{section_constraints}. Here density is far below the jamming transition point \cite{Nagel&Liu} and is not a control parameter of the system. 
One focuses on time scales much shorter than those typical of bubble coarsening and coalescence, so the \revision{$\mathcal{N}$ bubbles that compose the foam} preserve their integrity and size. Strictly, the area $A_{i}$ and pressure $\Pi_i$ of any bubble $i$ depend on the specific arrangement of all the bubbles, and so their values fluctuate in a shuffled foam; only the amount of gas that it contains is conserved. However, as discussed in Appendix A, the area and pressure fluctuations with foam configurations are usually very small \cite{livre_mousse} and so $A_i$ can be equated with its average value: $A_i\simeq\overline A_i$, where the upper bar denotes a \textit{time average} or, assuming ergodicity, an averaging over the foam configurations (\textit{ensemble average}). Hence, each bubble can be unambiguously identified with its area rather than the amount of gas it contains. To simplify the expressions hereafter, it is convenient to introduce the effective radius $R_i= \sqrt{A_i/\pi}$, \textit{i.e.} the radius of the circle which has the same surface area.

There are experimental and numerical evidence \cite{Quilliet} that the geometrical (size) and topological (number of sides) features of a bubble are correlated: in a 2D monodisperse foam (one bubble size), most of the bubbles are $6$-sided. Conversely, in a polydisperse foam, the larger bubbles have statistically more sides than the smaller ones. Correlations between geometrical and topological quantities of a foam are characterized by the conditional probability $P(n \vert R)$ for a bubble to have $n$ sides, given its size $R$ (conditional side number distribution). \revision{One assumes here discrete distributions of bubble sizes, although transposition of the equations to continuous distributions is straightforward.} The model predicts \cite{Durand1}:
\begin{equation}
P(n \vert R)=\frac{1}{\xi\left( R \right)}   \exp \left[ {-\frac{\beta \sqrt{\pi}}{3\epsilon R}n(n-6) +\mu n}\right],
\label{historic}
\end{equation}
where $\xi(R)=\sum_{n \geq n_{\mathrm{min}}} \exp \left[ {-\frac{\beta \sqrt{\pi} }{3\epsilon R} n(n-6) +\mu n}\right]$ is the partition function of the bubble. $n_\mathrm{min}$ is the minimal number of sides of a bubble, often taken as equal to $3$ in the literature, but in principle $n_{\mathrm{min}}=2$ is not forbidden by the space-filling constraints.
% $n_\mathrm{min}=1$ must be disregarded as it cannot be a metastable state). 
There is no free parameter in the model: $\epsilon\simeq 3.72$, while $\beta$ and $\mu$ are implicitly related to the (known) size distribution $P(R)$ through: $\partial \ln \Xi / \partial \beta=0$
and $\partial \ln \Xi / \partial \mu=6 \mathcal{N}$, 
where $\Xi$ is the partition function of the entire foam, defined through
\begin{equation}
\revision{\ln \Xi=\mathcal{N} \sum_{\lbrace R \rbrace}P(R) \ln \xi(R).}
%\ln \Xi=\mathcal{N} \int_{0}^{\infty}P(R) \ln \xi(R) \mathrm{d}R.
\end{equation}
No assumption is made on the shape of the size distribution $P(R)$; it can be unimodal or not. For a given $P(R)$, the equations of the model can be solved numerically. 
%This  (implicitly) relates the distribution of sides $p(n)$ to the distribution of sizes $p(A)$: $p(n)=\int_0^\infty p(A)p_A(n)\mathrm{d}A$.
The model predicts an order-disorder transition at a small, but finite, size dispersity \cite{Durand3}. Below this transition point (crystallisation threshold), all bubbles are 6-sided cells: the foam is topologically ordered. For a bidisperse foam, the transition occurs at the large-to-small size ratio $7/5=1.4$.
Above this point,
%Furthermore, 
analytical approximations can be derived: the truncated and discrete normal distribution (\ref{historic}) is evaluated by treating $n$ as a continuous variable \cite{Durand2,Durand3}. This approximation yields analytic expressions for $\beta$ \revision{and} $\mu$:
\begin{eqnarray}
\beta^{-1}=\dfrac{6 \sqrt{\pi}}{\epsilon}\left( \langle R^{-1} \rangle -\langle R \rangle^{-1} \right),
\label{beta}\\
\mu^{-1}=3\left( \langle R \rangle \langle R^{-1} \rangle -1 \right),
\label{mu}
\end{eqnarray}
Here the brackets denote an averaging over the bubbles that compose the foam: for any quantity $x_i$ attached to every bubble $i$, $\langle x \rangle=\sum_{i=1}^\mathcal{N} x_i/\mathcal{N}$. Introducing $p(R,n)$, the proportion of bubbles with size $R$ and number of sides $n$, this average can also be expressed as $\left< x \right>=\sum_{\lbrace R,n \rbrace}p(R,n)x(R,n)$. It should not be confused with the time (or ensemble) average defined above. 

Eqs. (\ref{beta})-(\ref{mu}) allow to express the average side number of a bubble with size $R$:
\begin{equation}
\overline{n}(R) \simeq 3\left(1+\dfrac{R}{\langle R \rangle} \right).
\label{nmoy(sqrtA)}
\end{equation}
The \textit{topological disorder} $ \sigma_n / \langle n \rangle =\sqrt{\langle n^2 \rangle - \langle n \rangle^2}/\langle n \rangle$ can also be expressed as a function of the (known) moments of the size distribution:
\begin{equation}
\left(\dfrac{\sigma_n}{\langle n \rangle}\right)^2=\frac{ 1 }{4}\left(\langle R \rangle \langle R^{-1} \rangle+\langle R^2 \rangle \langle R \rangle^{-2} -2 \right).
\label{topo-geo}
\end{equation}
In principle, the analytic expressions (\ref{beta}),(\ref{mu})(\ref{nmoy(sqrtA)}),(\ref{topo-geo}) are valid only for sufficiently high polydispersity of size. In practice, these are very good approximations down to the crystallisation threshold \cite{Durand3}.
 
Predictions on individual [Eq. (\ref{nmoy(sqrtA)})] and global [Eq. (\ref{topo-geo})] topology-geometry correlations agree well with experimental and numerical data of foams that are shuffled, either thermally \revision{(with Potts simulations)} or mechanically \cite{Durand2,Durand3}. Nevertheless, because of the very different nature of these two shuffling mechanisms, one does not expect to observe such agreement for higher shear rate or higher temperature, nor for all statistical properties of the foam. In particular, one expects the two shuffling mechanisms present different barrier-crossing statistics \cite{Barrat1,Barrat2}.

\section{Space-filling constraints in a foam at rest}\label{section_constraints}

A 2D foam is a partition of the plane without any gaps or overlaps, and its structure must obey a set of well established physical, geometrical, and topological constraints \cite{WeaireBook1,livre_mousse}.
The physical constraints follow from the mechanical equilibrium throughout the system: 
the balance of film tensions at every vertex yields Plateau's laws: the edges, or Plateau borders, meet in threes at a vertex, at an angle of $120^{\circ}$ with each other.
The balance of normal forces across each film yields Young-Laplace's law: every edge is an arc of circle, whose algebraic curvature $\kappa_{ij}$ is proportional to the pressure difference between the adjacent bubbles $i$ and $j$:
\begin{equation}
\kappa_{ij}=-\kappa_{ji}=\dfrac{\Pi_j-\Pi_i}{\gamma},
\label{Laplace}
\end{equation}
where $\gamma$ is the 2D film tension,
and $\Pi_i$ and $\Pi_j$ are the 2D pressures in bubble $i$ and $j$, respectively (by convention, $\kappa_{ij} \geq 0$ when the center of curvature is outside the cell $i$, \textit{i.e.}: when $\Pi_j \geq \Pi_i$). It must be emphasized that at microscopic scale the interface is not smooth, because of thermal fluctuations. In fact, Young-Laplace's law relates the \textit{mean} curvature to the \textit{mean} bubble pressures.

There is also a constraint on every bubble because of the fixed amount of gas that it contains. This constraint is often replaced, for simplicity (see Appendix A), by a constraint on every bubble area (incompressible gas).

Euler's rule is a topological constraint which relates the number of bubbles $\mathcal{N}$, with those of Plateau borders $N_{Pb}$, and vertices $N_v$ in a foam that covers all the 2D space: 
\begin{equation}
\mathcal{N}-N_{Pb}+N_v=\chi,
\label{Euler}
\end{equation}
where $\chi$ is the Euler-Poincar\'{e} characteristic (\textit{e.g.} $\chi=0$ for a toroidal space, $\chi=2$ for a spherical one). \revision{This formula can also be used for a foam that partially covers the 2D space (free cluster), by considering the rest of the space as an extra bubble -- provided that it is homeomorphic to a disk \cite{Euler}.}

Finally, the Gauss-Bonnet theorem relates geometry and topology of every bubble $i$: 
\begin{equation}
\sum_{j \in \mathfrak{N}(i)} l_{ij} \kappa_{ij}=\frac{\pi}{3} (n_i-6),
\label{Gauss-Bonnet}
\end{equation}
where $\mathfrak{N}(i)$ represents the set of the $n_i$ bubbles that neighbor bubble $i$, and $l_{ij}$ is the length of the edge common to bubbles $i$ and $j$. 

The constraints listed above substitute for the steric repulsion (excluded volume) that lies in hard-core granular materials: as a consequence, there is no limit on the number of neighbours that a bubble can have (while in a monodisperse disk packing, no more than $6$ neighbours can fit around a disk because of steric exclusion). Curvature and length of the sides can adapt to satisfy Young-Laplace's and Plateau's laws; this induces an increase of pressure difference between the central bubble and its neighbours.
Geometry and topology of the bubbles are then correlated through these space-filling constraints.

\section{Shuffled foams}
\label{section_shuffling}

\subsection{Mechanical and thermal shuffling of foams}

From the point of view of standard thermodynamics, a liquid foam at rest is an out-of-equilibrium system: it has many metastable states, but the energy required to jump from one metastable state to another is much higher than the thermal energy $k_BT$ within a usual range of temperature. Thus, in absence of coarsening and coalescence, it takes an indefinitely long time to escape from an energetic valley. Actually, the temperature required to trigger a T1 event is far above the vaporizing temperature of the liquid phase, which makes thermal shuffling inaccessible experimentally.
%(maybe low surface tension polymer colloid substances ? citer Lekerkerker). 
Numerically, on the other hand, this mode of shuffling is easy to handle. This can be done with Monte Carlo simulations (see Appendix B). \revision{Such a system is at thermodynamic equilibrium, and the probability of being in a given configuration follows the Boltzmann distribution.}

Mechanical shuffling also confers dynamics to the foam and can be achieved both numerically and experimentally. However, such a system is still out of thermodynamic equilibrium: energy is injected (in the form of mechanical work) to the system by a non-thermal source, which is then dissipated during the rearrangements associated with T1 events \cite{Durand&Stone}.

\subsection{Gentle shuffling of foams}

Foams can explore their phase space by various shuffling mechanisms. One restricts this study to \textit{gentle shufflings}, defined as follows.

\revision{For thermally shuffled foams, the range of temperature used in simulations is $\Theta \lesssim\gamma\langle R \rangle$: enough to trigger T1s and shuffle the foam but too small to eject bubbles from a cluster. 
It must be stressed that, although T1 events are rare in this low temperature regime, equilibration is guaranteed (but takes a very long time to be achieved) for any temperature larger than $T=0$. Different numerical techniques (\textit{e.g.} \cite{Hukushima}) allow to speed up tremendously the evolution towards equilibrium.}

For mechanically shuffled foams, one assumes that shuffling is very slow when compared to the typical relaxation time associated with a T1 event: the transient states can then be neglected in the ensemble of accessible states: foam passes instantaneously from one metastable state to another. Moreover, one considers only shuffling mechanisms where the bubble areas, and hence the total foam area, are preserved. Because of the high deformability of its constituents, such a restriction is always possible for a foam, unlike for jammed hard granular systems \cite{Bowles}. %Continuity of the trajectories in the phase space is then ensured \cite{Bowles}.
One also supposes that T1 events occur homogeneously within the foam. This assumption certainly restrains the mode of shuffling: \revision{for instance, simple shear (with displacements imposed on the top and bottom boundaries only \cite{note_uniform_strain}) may not be a good shuffling process in this perspective, as some studies have reported a localization of the T1s} near the box boundaries (shear banding phenomenon) \cite{Dennin,shear-banding}. Conversely, a random, non periodic, shuffling should satisfy this hypothesis. It seems that such an experiment has never been done yet.

Thus, for both shuffling mechanisms, the structure of the foam evolves though T1 events exclusively, and bubble preserve their integrity.

\subsection{Equiprobability hypothesis}\label{section_postulate}

By analogy with the fundamental postulate of statistical mechanics \cite{Books}, one assumes that, under a gentle and homogeneous shuffling, all the accessible microstates of a large foam ($\mathcal{N}\gg 1$) are visited with equal probability.

One may wonder whether or not the equiprobability hypothesis must be restricted to the states with same energy, as it is postulated for thermal systems at equilibrium.
Actually, the answer depends on the origin of the shuffling.

For foams which are mechanically shuffled, the answer to this question is still a large field of investigation: as previously mentioned, such system is out-of-equilibrium from the point of view of conventional thermodynamics \cite{Edwards-and-co,Blumenfeld,Barrat1,Barrat2}.

\begin{figure}
\begin{center}
\includegraphics[width=\columnwidth]{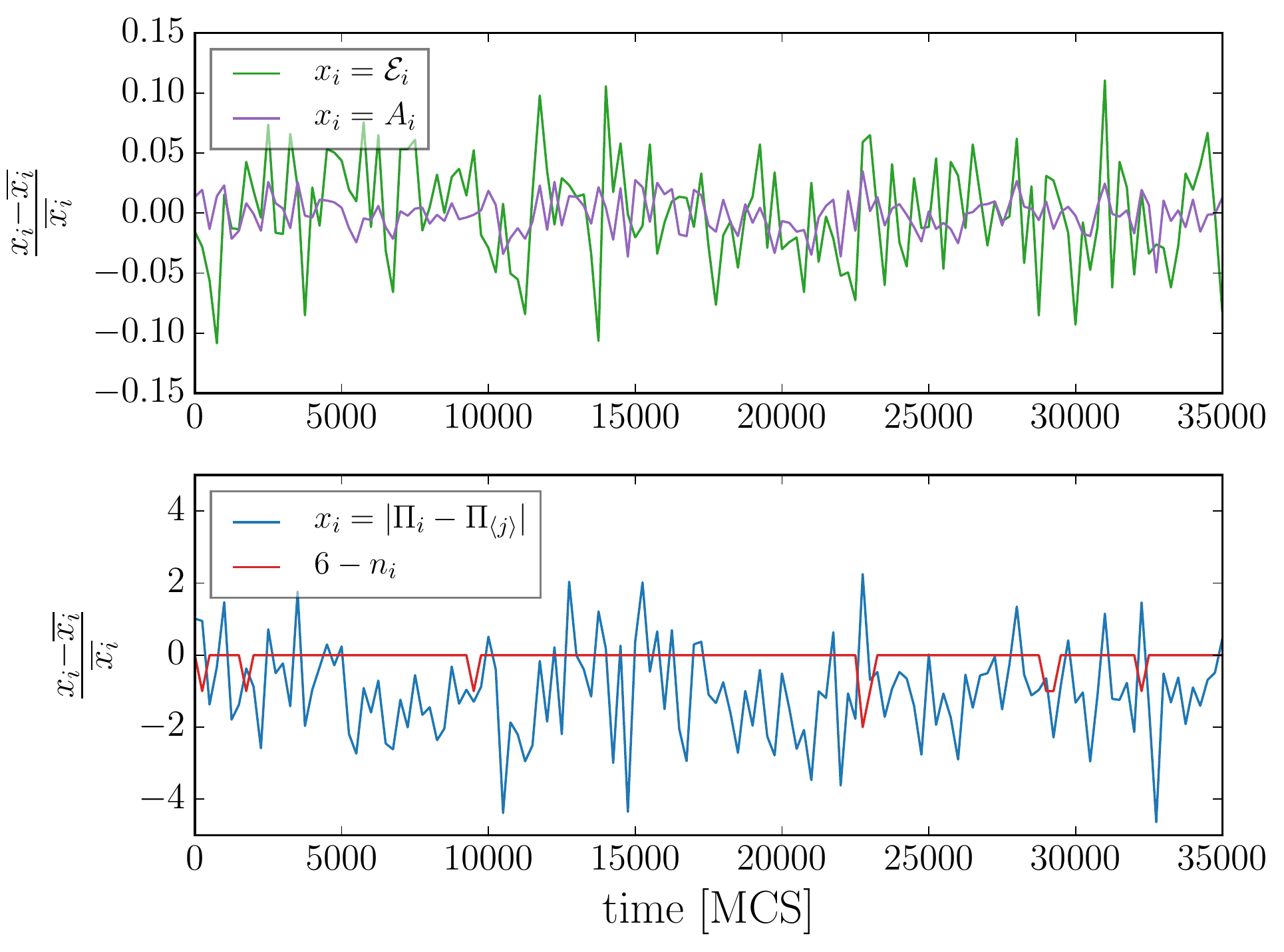}
\caption{Typical \revision{relative} fluctuations for different quantities attached to bubble $i$: area $A_i$ (purple curve), energy $\mathcal{E}_i$ (green curve), and pressure difference $\vert\Pi_i-\Pi_{\langle j \rangle}\vert$ (blue curve), where $\Pi_{\langle j \rangle}=\sum_{j \in \mathfrak{N} (i)}\Pi_j/n_i$. Curves are shown on two different graphs because of the difference in amplitude of the \revision{relative} fluctuations. Bubble perimeter is not shown, as it is barely distinguishable from the energy curve. Topological charge $q_t(n_i)=6-n_i$ (red curve) is also drawn to indicate the T1 events in which bubble $i$ is involved. 
}
\label{cell-curvature-vs-perimeter}
\end{center}
\end{figure}

\revision{
For thermally shuffled foams, equiprobability should be restricted to states with same energy. However, in the gentle shuffling regime considered here, the low temperature range is such that variations of energy between the visited states can be neglected.}
Fig. \ref{cell-curvature-vs-perimeter} shows the \revision{relative} fluctuations observed for different quantities attached to a single bubble $i$ within a thermally shuffled foam, using Cellular Potts Model simulations (see Appendix B). As expected (see Appendix A), variations of the bubble area are much smaller than those of the pressure difference $\vert\Pi_i-\Pi_{\langle j \rangle}\vert$, where $\Pi_{\langle j \rangle}=\sum_{j \in \mathfrak{N}(i)} \Pi_j / n_i$ is the \textit{instantaneous} average pressure in the bubbles that belong to $\mathfrak{N}(i)$, the neighbourhood of bubble $i$.
\revision{Taking the absolute value of the pressure difference ensures strictly positive averaged value.}

Energy fluctuations are also very small. Almost no energy is exchanged between bubbles, which justifies that the Boltzmann weighting can be neglected in a gentle thermal shuffling. This is also consistent with previous studies that show that the difference of energy between metastable states are less than 2\% for a 2D foam \cite{Jiang,Graner1}. In the mean-field approximation detailed in Sect. \ref{section_MFA}, bubbles have constant perimeters: the energy of every bubble, and hence of the entire foam, are fixed. 

For these reasons, one can safely disregard the energy weighting of the probability distribution, and not distinguish thermally from mechanically shuffled foams.
Indeed, in spite of the profound difference of these two modes of shuffling, experimental and numerical data collapse extremely well \cite{Durand2,Durand3} in the gentle shuffling regime.
\revision{
To summarize, in the gentle shuffling regime the energy plays a role in the selection of the accessible states (since they correspond to the local minima of energy), but not in the probability distribution over all these accessible states.}

\revision{
Because the dynamics must obey to a local conservation law on the number of sides of the four bubbles involved in a T1 event, it is not granted that the ensemble of the microstates visited by a succession of T1 processes is equal to the ensemble of states which are compatible with the space-filling constraints listed in Sect. \ref{section_constraints}. In other terms, it is not obvious that the foam can pass from any configuration that satisfies these space-filling constraints to any other one through a sequence of T1 events. The resolution is specially not trivial for an \textit{infinite} foam (\textit{i.e.}, with no boundaries). However, for a real, bounded foam, the total number of sides is allowed to fluctuate (see Sect. \ref{section_self}), and topological charges can appear or disappear easily during T1s at the boundary. For that reason, it can be reasonably conjectured that the two ensembles are identical in that case.
}

\subsection{Degrees of freedom of a gently shuffled foam}\label{section_dof}
To perform a statistical description of the foam structural properties, one needs to enumerate the microstates which correspond to the same macroscopic state. 
A detailed description giving positions and momenta at the atomic scale is too cumbersome to handle, and one must use instead a coarse-grained description in which the state of the foam refers to its structure (topology and geometry) at the bubble scale.
For instance, the complete knowledge of the contour of every bubble
fully describes a metastable state \cite{Nelson}. 
Actually, contours are not independent in a foam at rest, because of the space-filling constraints. The minimal number of degrees of freedom that are required to describe a foam metastable state is still an open question \cite{Graner1,Fortes1,Weaire2,Vaz3,Moukarzel}. Probably the most comprehensive study on this domain has been carried out by C. Moukarzel which has identified a set of $4\mathcal{N}$ independent variables to fully describe any metastable configuration of a foam that tiles the plane \cite{Moukarzel}. However, these variables have no direct physical interpretation. %Moreover, these variables are not independent anymore when the bubble areas/gas contents are prescribed. 
Here instead, one assumes that a microstate $L$ of the foam is correctly
%properly/accurately /adequately
described by the number of sides $n_i$, \revision{centroid} location $\mathbf{r}_i$ and pressure $\Pi_{i}$ of every bubble: $L\equiv\lbrace n_i, \mathbf{r}_i, \Pi_{i};~i=1,2,\dots,\mathcal{N}\rbrace$. Although pressure fluctuations are small, the fluctuation of pressure difference between adjacent bubbles can be large (see Appendix A). That is why they must be taken into account in the description of a microstate. 

\revision{With this choice of $4\mathcal{N}$ variables}, the microstate of the foam $L$ is given by the microstate $\ell_i=(\mathbf{r}_i,n_i,\Pi_i)$ of every bubble $i$, which is convenient to apply a mean field description.
However, this choice is not crucial for the rest of the paper:
the statistical description presented here only requires that a microstate of the foam is adequately described by the variables $\lbrace n_i \rbrace$, plus possible other degrees of freedom which must be independent of the $\lbrace n_i \rbrace$.

When a bubble $i$ is directly involved in a T1, its associated variables $(\mathbf{r}_i,n_i,\Pi_i)$ vary together, inducing correlations between them. However, bubble positions $\lbrace\br_k \rbrace$ are continuous variables and have some uncertainty attached to them \cite{note_fluctuations}. 
Thus, after few T1 events, $\mathbf{r}_i$ and $(n_i,\Pi_i)$ can be fairly treated as independent variables: at a given value of $(n_i,\Pi_i)$ correspond many different values of $\mathbf{r}_i$, and \textit{vice-versa}. Note that the same kind of assumption is done in the statistical mechanics of a gas (\textit{Stosszahlansatz} or \textit{molecular chaos hypothesis} \cite{Books}) to justify the independence of positions and momenta of the particles.
In addition, thanks to the high deformability of the bubbles, $(n_i,\Pi_i)$ are almost independent of the relative positions of the $\mathcal{N}-1$ other bubbles $\lbrace\mathbf{r}_j-\mathbf{r}_i ,~j\neq i\rbrace$. Therefore, the variables $\lbrace\mathbf{r}_i \rbrace$ and $\lbrace n_i,\Pi_i \rbrace$ can be treated as two sets of variables that are independent from each other.

It is shown in the next Section that, within a mean field description, variables $\lbrace\mathbf{r}_i \rbrace$ and $\lbrace n_i \rbrace$ are sufficient to properly describe a foam microstate.

\section{Mean Field Approximation} \label{section_MFA}

Because of the space-filling constraints (see Sect. \ref{section_constraints}), bubbles are correlated objects : a change in the geometry or topology of a bubble affects the geometry and topology of the other bubbles.
Therefore, the number of accessible states cannot be simply obtained by enumerating the possible distributions of the $N$ sides over the $\mathcal{N}$ bubbles: many of these distributions are not compatible with the space-filling constraints. A perfect description taking all these constraints into account is illusive. 
Instead, a mean field approach is proposed, in which the fluctuations of the neighbourhood of a bubble with given size $R$ and given number of sides $n$ are neglected. To legitimate this approximation, one shows below that in most cases these correlations are in fact short-ranged.

\subsection{Screening of bubble correlations} \label{correlations}

One discusses how the bubble microstates $\ell_i \equiv (n_i,\mathbf{r}_i,\Pi_i)$ are correlated to each other. One should first study how they are altered by a unique T1 event in the foam: clearly, only the four bubbles involved in the T1 have their $n_i$ that vary. The pressure field is \textit{a priori} modified on a longer range: previous studies \cite{Graner1,Cox2} have shown that the extra pressure due to a \textit{topological charge} $q_t(n)$ (a bubble with $n \neq 6$ sides) within a regular foam of hexagonal bubbles decreases logarithmically (as for the electric potential generated by a point charge in a 2D dielectric medium). However, a large foam must contain in same proportion positive ($q_t=6-n>0$) and negative ($q_t<0$) topological charges. For instance, a T1 event in a hexagonal foam produces two positive and two negative topological charges simultaneously. Their antagonist effects on the pressure field screen their correlations on a typical lengthscale of one bubble size. This ``screening'' effect is further enhanced in a disordered foam \cite{Cox2}.

\begin{figure}
\begin{center}
\subfigure[]{
\includegraphics[width=0.23 \textwidth]{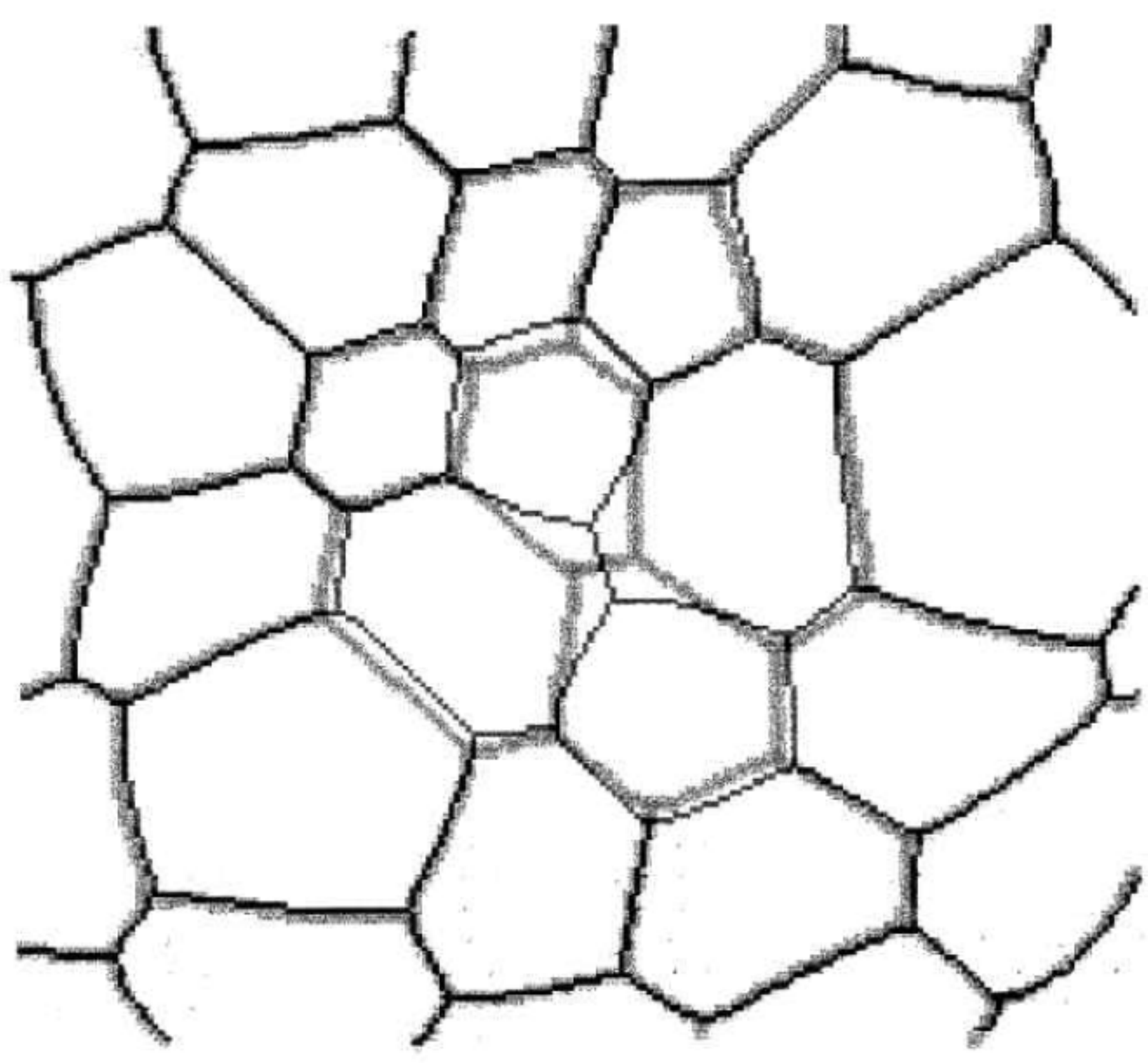}}
%\caption{T1-superposition Elias et al. Excerpt from \cite{Elias}}
\subfigure[]{
\includegraphics[width=0.22 \textwidth]{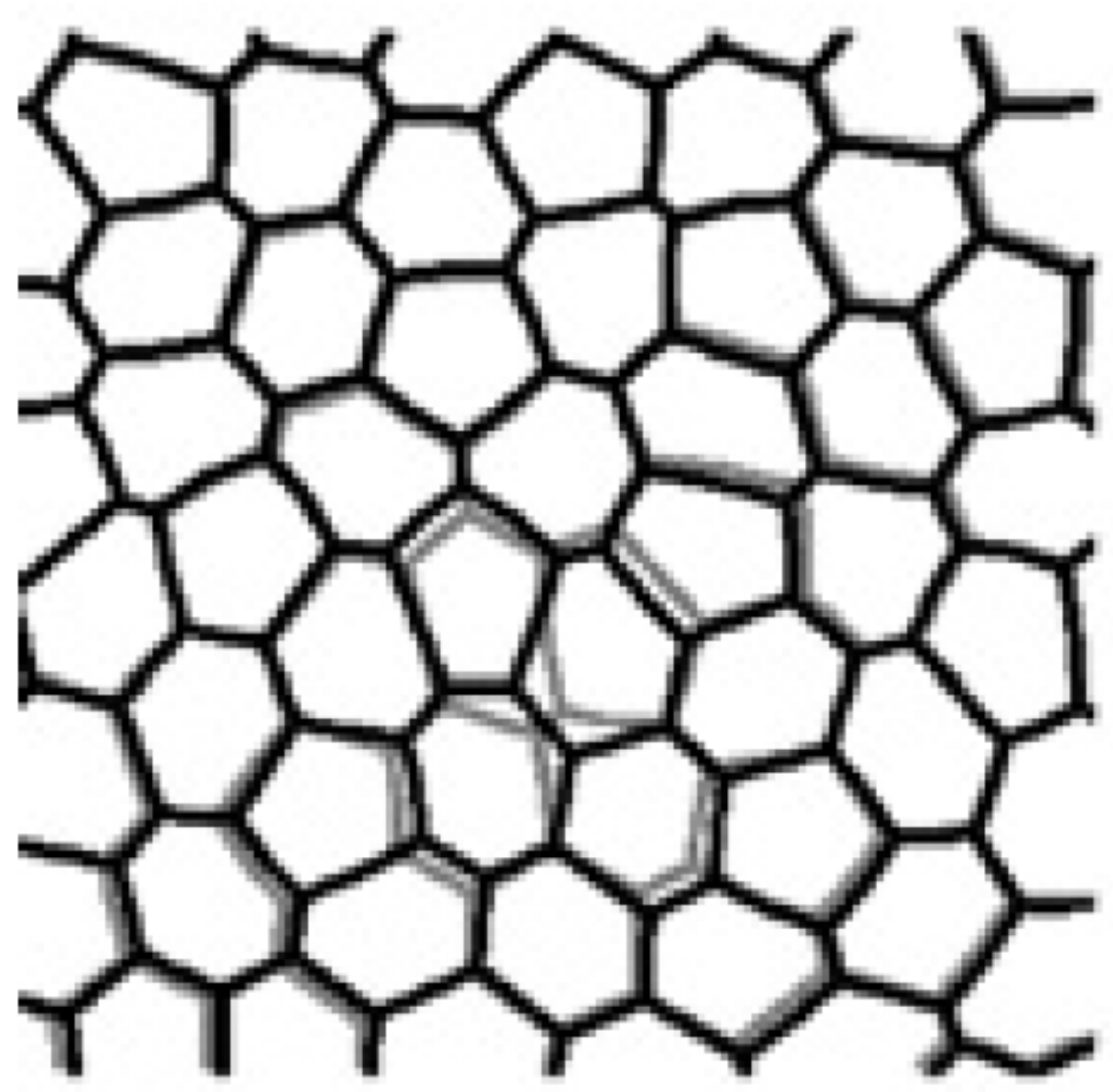}}
\caption{Superposition of the metastable states of a foam before and after a T1 event; (a) experimental foam, excerpt from \cite{Elias}; (b) numerical foam (Surface Evolver), excerpt from \cite{Cox2}.}
%\caption{T1-superposition Cox et al}
\label{T1-superposition}
\end{center}
\end{figure}

Displacements are also very localized: Figure \ref{T1-superposition} shows the superposition of the pictures of a dry foam before and after a T1 event, either experimentally or numerically.
Essentially the four bubbles directly involved in the T1 event have their positions and geometries modified; the rest of the foam structure is unperturbed. 
Thus, as long as the bubble $i$ is not involved in a T1 event, it remains almost in the same microstate $\ell_i$: its number of sides $n_i$ is fixed, and the change of its position $\br_i$ and pressure $\Pi_i$ are imperceptible. Precisely, the fluctuations of position $\delta \br_i$ due to T1s elsewhere in the rest of the foam are small \revision{compared to the lengths of the bubble sides.}
Therefore, as long as T1 events remain diluted, correlations between bubble microstates $\ell_i$ are short-ranged: the variables $(n_i,\mathbf{r}_i,\Pi_i)$ of a bubble are primarily correlated to those of its first neighbours. It can be noticed that correlations of bubble displacements are also short-ranged. This situation is very different from what is observed in dense packings of hardcore particles: because of steric exclusion, the positions of the particles are not independent, yielding macroscopic correlation length of the displacement. Direct measures of bubble displacements in partially wet foams (liquid fraction $\simeq 10 \%$) confirm the short correlation length of bubble displacements, typically of a few bubble radii \cite{Durian1991,Duri2009}, and hence much lower than what is observed in dense granular packings of hardcore particles \cite{Duri2009}.

\subsection{Isotropic (regular) bubble model} \label{IPP}
\begin{figure}[htb]
\begin{center}
\subfigure[]{
\includegraphics[width=0.225 \textwidth]{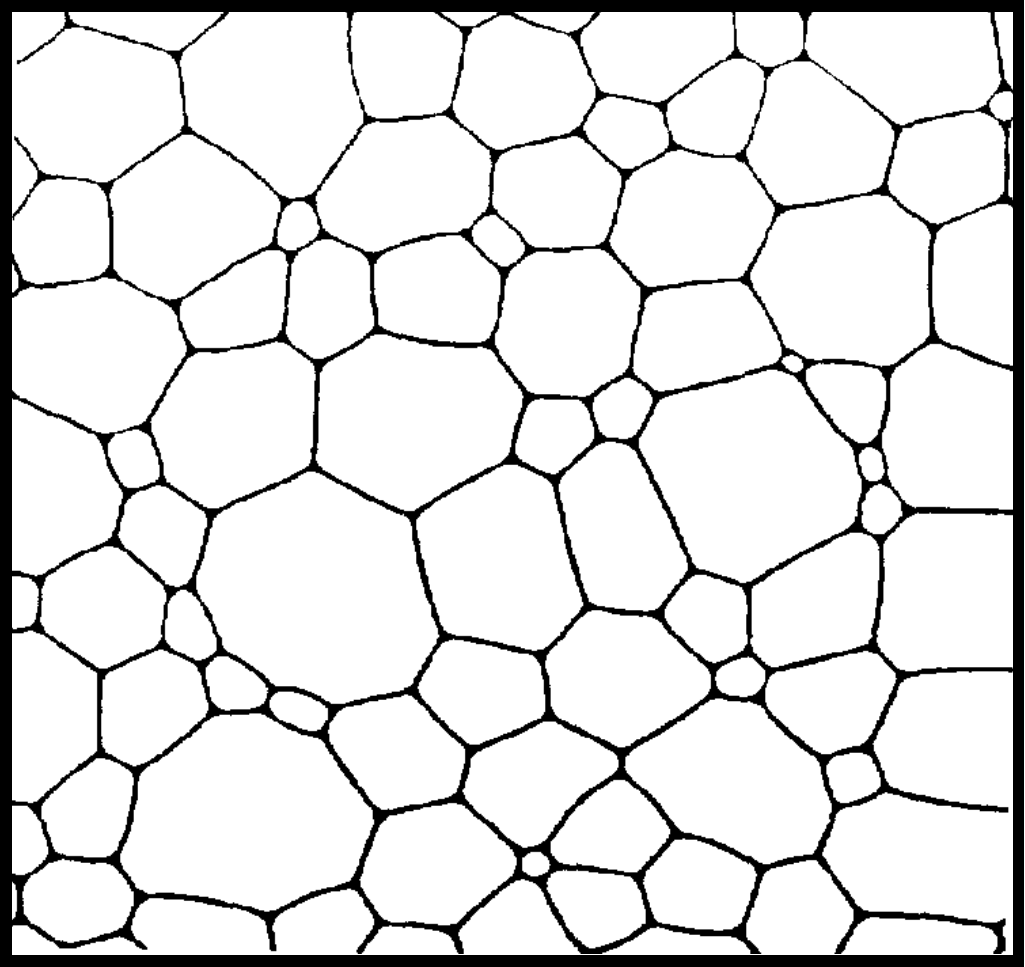}
}
\subfigure[]{
\includegraphics[width=0.225 \textwidth]{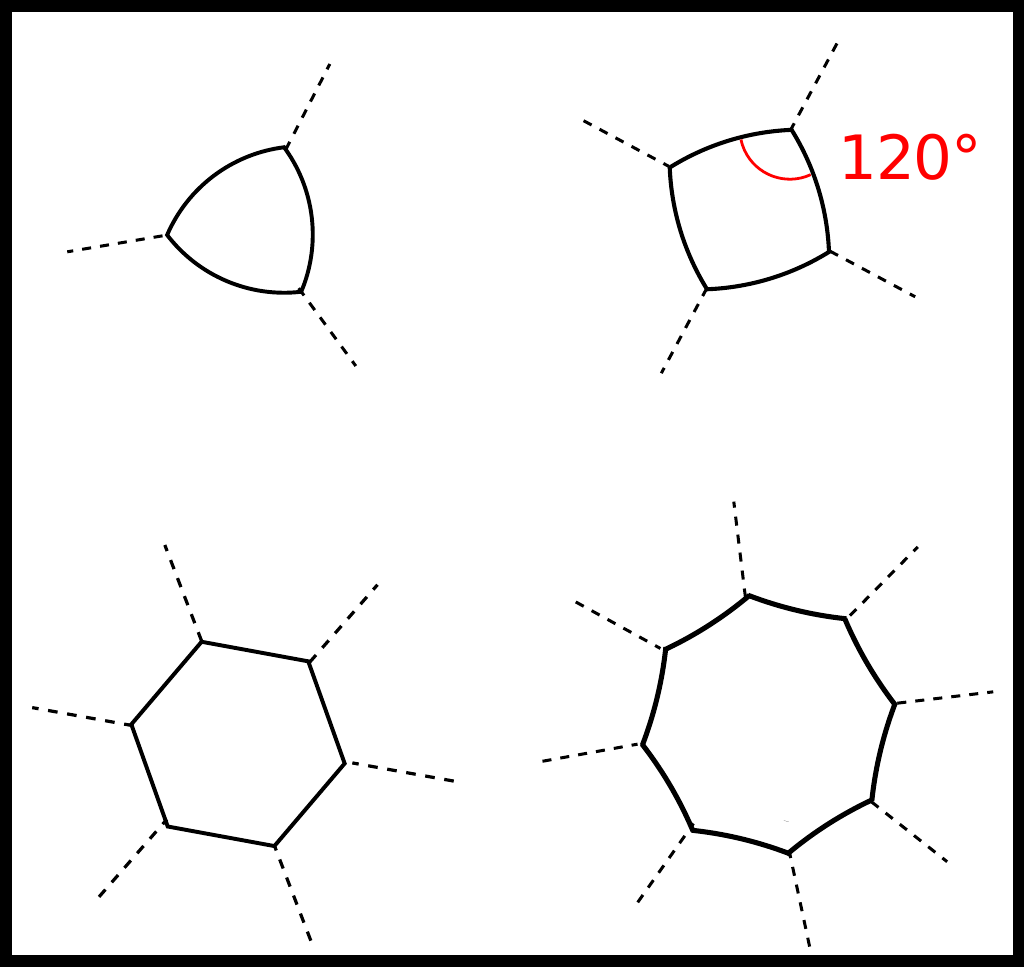}
}
\caption{(a): Real two-dimensional foam in one of its metastable states. (b): Mean-field approximation: each bubble $i$ is a regular cell surrounded by a uniform and isotropic foam, resulting form the averaging of the positions and side numbers of the $\mathcal{N}-1$ other bubbles.}
\label{2dfoams}%
\end{center}
\end{figure}

The proposed Mean Field Approximation (MFA) consists in neglecting the fluctuations of the neighbourhood of each bubble with given size $R$ and number of sides $n$. In other terms, the geometry of such a bubble is equated with the one obtained from averaging over all the foam configurations for which the number of sides of this bubble is $n$. 
Since all the constraints listed in Sect. \ref{section_constraints} are satisfied for every configuration $L$ of the foam, one looks for an average bubble geometry that satisfies these constraints too. Assuming that the foam is homogeneous and isotropic, the bubble is surrounded by a uniform foam (identical bubbles): all sides are equivalent and the geometry must correspond to the geometry of a regular $n$-sided cell, for which sides are identical in shape and length (see Fig. \ref{2dfoams}). 

The geometry of such regular cell is entirely determined by its size and number of sides \cite{Graner1,Iglesias}. In particular, the curvature of each side is $\sqrt{\pi}(n-6)/\left(3\epsilon(n)R\right)$ (by convention, the curvature is chosen positive when its center is outside the bubble).
$\epsilon(n)$ is the elongation of the cell (ratio of perimeter to square-root of area). 
It is a very slowly decreasing function, lying between $\epsilon(2)\simeq 3.78$ and $\epsilon(\infty)\simeq 3.71$. In fact, 2-sided cells are seldom observed in gently shuffled foams, and $\epsilon(3)$ already drops down to $\simeq 3.74$. Exact expression of $\epsilon(n)$ is given in refs \cite{Durand1,Graner1,Iglesias}. Hereafter, this weak dependence is neglected, and one notes  $\epsilon=\epsilon(6)\simeq  3.72$. It can be noticed that this approximation implies that the perimeter of a bubble does not depend on its number of sides; hence, the energy of every bubble, and thus of the entire foam, is fixed. The restriction of the equiprobability hypothesis to the configurations of same energy discussed in Sect. \ref{section_postulate} is superfluous within this MFA.

The regular bubble (rb) model presented here assumes that the difference of pressure between a bubble $i$ and its neighbourhood depends primarily on the size and number of sides of that bubble, and not on the size polydispersity of the foam. This may fail for very large size-dispersity values. 

It will be useful to introduce also the \textit{bubble} curvature of a $n-$sided bubble, defined as the sum of the algebraic curvatures of its sides. For the regular bubble, its expression is:
\begin{equation}
\kappa_{\text{rb}}(R,n)= \frac{\sqrt{\pi}}{3}\frac{n\left(n-6 \right)}{\epsilon R}.
\label{regular_bubble_curvature}
\end{equation}
The \textit{side} curvature of such a bubble is just $\kappa_{\text{rb}}(R,n)/n$.

\subsection{Self-consistency equations}\label{section_self}
By construction, a regular bubble satisfies most of the constraints listed in Sect. \ref{section_constraints}. Only two sorts of constraints are not automatically satisfied in an idealized foam where all the bubbles have regular shapes: a global constraint caused by the Euler's formula [Eq. (\ref{Euler})], and local constraints caused by the Young-Laplace's law [Eq. (\ref{Laplace})] at every Plateau border. In the following, \revision{one formally shows that the MFA is equivalent to replace these local constraints by a second global constraint.}
Consider first the rather academic case of a foam with no boundary, \textit{i.e.} that covers entirely the 2D space.
Since all the vertices of such a foam are 3-connected, Eq. (\ref{Euler}) yields the first global constraint:
\begin{equation}
\sum_{i=1}^\mathcal{N} n_i=6(\mathcal{N}-\chi).
\label{self-consistency-eq1}
\end{equation}
To derive the second global constraint, it can be first noticed that since a bubble with $n_i$ sides belongs to the neighbourhood of $n_i$ different bubbles in an unbounded foam, pressures in adjacent bubbles must obey the following identity:
\begin{equation}
\revision{
\sum_{i=1}^\mathcal{N} n_i^L \Pi_{i}^L=\sum_{i=1}^\mathcal{N} \sum_{j \in \mathfrak{N}^L(i)}\Pi_{j}^L.}
%\sum_{i=1}^\mathcal{N} n_i \left( \Pi_i-\Pi_{\langle j \rangle(i)} \right)=0.
\label{total_curvature}
\end{equation}
\revision{Superscripts $L$ have been momentarily added to the fluctuating variables to emphasize their dependence on the specific foam configuration $L$. Note that the set of $n_i^L$ bubbles that constitute the neighborhood $\mathfrak{N}^L(i)$ of bubble $i$ changes with configurations. It can also be noticed that the identity (\ref{total_curvature}) is not specific to bubble pressures, but also holds for any other quantity $x_i^L$ that is defined on every bubble $i$.
}

Equations (\ref{self-consistency-eq1}) and (\ref{total_curvature}) are satisfied for every configuration $L$ of the foam. Therefore they must also be true on average. Let $p_L$ be the probability for the foam to be in configuration $L$, and $p_i(n)=\revision{\sum_{\lbrace L \vert n_i^L=n \rbrace}p_L}$, the probability that bubble $i$ has $n$ sides \revision{(the sum being over all foam microstates $L$ which satisfy $n_i^L=n$). The pressure difference between bubble $i$ and its neighbours, averaged over all configurations for which $n_i^L=n$, is } 
\begin{equation}
\revision{
\widehat{\Delta \Pi}_i(\lbrace R_k \rbrace,n)= \sum_{\lbrace L \vert n_i^L=n \rbrace}p_L(\Pi_i^L-\Pi_{\langle j \rangle}^L)/p_i(n),
}
\label{Eq13}
\end{equation}
\revision{
where $\Pi_{\langle j \rangle}^L=\sum_{j \in \mathfrak{N}^L(i)} \Pi_j / n_i$ is the instantaneous average pressure in the bubbles that belong to $\mathfrak{N}^L(i)$ This \textit{partial} averaging is denoted by the hat symbol $\widehat{~}$. As this neighborhood changes with configuration $L$, $\widehat{\Delta \Pi}_i$ is \textit{a priori} a function of $n$ and all the bubble sizes $\lbrace R_k;k=1,\dots, \mathcal{N}\rbrace$.}

Averaging of Eqs (\ref{self-consistency-eq1}) and (\ref{total_curvature}) over all foam configurations $\lbrace L \rbrace$ leads to:
\begin{align}
\sum_{i=1}^\mathcal{N} \sum_{n\geq n_{\mathrm{min}}}p_i(n) n & = N, \label{total_side_number} \\
\revision{
\sum_{i=1}^\mathcal{N} \sum_{n\geq n_{\mathrm{min}}}p_i(n) n\widehat{\Delta \Pi}_i(\lbrace R_k \rbrace,n) } & = 0,
\label{total_curvature_bis}
\end{align}
where $N=6(\mathcal{N}-\chi)$ is the sum of side numbers. On the other hand, the MFA presented above consists in neglecting the fluctuations of the neighbourhood for every bubble $i$ of given number of sides $n_i$:
\begin{equation}
\revision{
\sum_{j \in \mathfrak{N}^L(i)} (\Pi_i^L-\Pi_j^L) \simeq n_i \widehat{\Delta \Pi}_i(\lbrace R_k \rbrace,n_i).
}
\end{equation}
Neglecting these fluctuations in Eq. (\ref{total_curvature}) yields:
\begin{equation}
\revision{
\sum_{i=1}^\mathcal{N} n_i \widehat{\Delta \Pi}_i(\lbrace R_k \rbrace,n_i) =0
%\label{total_curvature}
}
\end{equation}
for every foam configuration. Introducing $P(n\vert R=R_i)$, the proportion of bubbles with size $R_i$ that have $n$ sides, this equation rewrites:
\begin{equation}
\sum_{i=1}^\mathcal{N} \sum_{n\geq n_{\mathrm{min}}} P(n\vert R=R_i) n \widehat{\Delta \Pi}_i(\lbrace R_k \rbrace,n) =0.
\label{total_curvature_ter}
\end{equation}
As explained in Sect. \ref{correlations}, correlation lengths are short (compared with the system size) for a foam far from the crystallization transition. By design, correlation length is zero in the MFA. Averaging over all the configurations of bubble $i$ (time or ensemble average) can then be equated with averaging over all the bubbles that have same size than bubble $i$ within the foam (spatial average), \textit{i.e.}: $p_i(n)=P(n\vert R=R_i)$. Then, Eqs (\ref{total_curvature_bis}) and (\ref{total_curvature_ter}) are identical.

This result shows that the MFA consists in replacing the local constraints [Eq. (\ref{Laplace})] by a unique global constraint [Eq. (\ref{total_curvature_ter})]. 
This reduction of the number of constraints is accompanied by a reduction of the number of degrees of freedom necessary to describe a foam microstate $L$, as the regular bubble model relates pressures $\lbrace\Pi_i\rbrace$ and side numbers $\lbrace n_i\rbrace$.

The mean bubble curvature is defined as 
\begin{equation}
\revision{
\widehat{\kappa}_i(\lbrace R_k \rbrace,n_i)=-n_i\widehat{\Delta \Pi}_i(\lbrace R_k \rbrace,n_i)/\gamma.
}
\end{equation}
Using the geometry of a regular bubble [Eq. (\ref{regular_bubble_curvature})], $\widehat{\kappa}_i(\lbrace R_k \rbrace,n_i)\equiv \kappa_\text{rb}(R_i,n_i)$,
and Eq. (\ref{total_curvature_bis}) becomes:
\begin{equation}
\sum_{i=1}^\mathcal{N} \sum_{n\geq n_{\mathrm{min}}}p_i(n) \kappa_\text{rb}(R_i,n)=0.
\label{self-consistency-eq2}
\end{equation}
For a foam, $\gamma$ is a constant that has been dropped out. However, for the extension of the model to other cellular systems, like epithelial tissues, interfacial tension can have different values \cite{PNAS}, and the regular bubble model must be adapted accordingly.

Equalities (\ref{total_side_number}) and (\ref{self-consistency-eq2}) hold for foams with no boundaries. 
They can be easily extended to the case of a free bubble cluster (\textit{i.e.} a foam that does not cover all the 2D space), or a set of individual bubbles within a larger foam. Assuming again $p_i(n)=P(n\vert R=R_i)$, one as:
\begin{align}
\sum_{i=1}^\mathcal{N} \sum_{n\geq n_{\mathrm{min}}}P(n\vert R=R_i) n=\N,
\label{constraints1}  \\
\sum_{i=1}^{\mathcal{N}} \sum_{n\geq n_{\mathrm{min}}}P(n\vert R=R_i) \revision{\kappa_\text{rb}(R_i,n)}=\K,
%\label{general_self_consistency_eqs}
\label{constraints2} 
\end{align}
\revision{where $\N$ and $\K$ are respectively the averaged values of the total number of sides $N^L=\sum_i n_i^L$ and total bubble curvature $K^L=\sum_i \sum_{j \in \mathfrak{N}^L(i)} \kappa_{ij}$ of the $\mathcal{N}$ bubbles that constitute the system. Averaging is performed over \textit{all} the microstates $L$ of the foam. }
Eqs (\ref{constraints1}) and (\ref{constraints2}) constitute self-consistency equations of the MFA. A foam with no boundary can be seen as the special case where $N=6(\mathcal{N}-\chi)$ and $K=0$. Otherwise, $N$ and $K$ are fluctuating variables. \revision{However, here and thereafter, superscripts $L$ are omitted to make the notation less cluttered}.
Clearly, $N$ and $K$ are independent variables: there are many ways of distributing $N$ sides to the $\mathcal{N}$ bubbles, leading to different values of $K$ (and \textit{vice-versa} when $\mathcal{N}\gg1$).

For a free cluster, adjacency of bubbles imposes that sides cancel in pairs, except for those at the boundary of the cluster. The curvature $K$ is then reduced to the sum of the curvatures of the sides that belong to the outer boundary $\mathcal{B}$: 
\begin{equation}
K=\sum_{i \in \mathcal{B}}\kappa_{i0}=\sum_{i \in \mathcal{B}}\dfrac{\Pi_0-\Pi_i}{\gamma},
\end{equation}
where $\Pi_0$ is the surrounding pressure. 
Values of $K$ and $N$ (and therefore of $\K$ and $\N$) are restricted within a certain range: $K<0$ since the pressure in any bubble of the cluster is higher than the external pressure \cite{Vaz3}. Combining Euler's rule [Eq. (\ref{Euler})] and Plateau's law gives $N_{Pb}=3(\mathcal{N}+1-\chi)$ for a cluster. Since two adjacent bubbles share the same Plateau border, each Plateau border in the bulk contributes to two bubble sides, while each Plateau border at the outer boundary contributes to only one bubble side. Thus, $ N<6(\mathcal{N}+1-\chi)$.

Self-consistency equations (\ref{constraints1}) and (\ref{constraints2}) are also valid for a foam that fills a container; in that case $N$ fluctuates, while $K=\K=0$. 
 
Fluctuations of $N$ and $K$ around their respective average values are characterized by $\delta N/\N$ and $\delta K/\K$, where $\delta X=\sqrt{\overline{X^2}-\overline{X}^2}$ is the standard deviation of the variable $X$, calculated over all foam configurations ($\delta X$ must not be confused with the standard deviation $\sigma_X$ calculated on bubble population). It is shown in Appendix A that these fluctuations become negligible as the number of bubbles $\mathcal{N}$ increases. However, $\delta N/\N$ and $\delta K/\K$ scale as $\mathcal{N}^{-\alpha}$ with exponents $\alpha>0$ that may differ from the value $1/2$ characteristic of additive variables \cite{Books}: indeed, $\alpha$ depends on the boundary conditions considered (free cluster, unbounded foam, or set of separate bubbles). Since fluctuations vanish in the limit $\mathcal{N}\rightarrow\infty$, the postulate of equiprobability presented in Sect. \ref{section_postulate} can be rephrased:
\textit{all the accessible states with same values of total curvature ($=\K$) and number of sides ($=\N$) are equally probable}.

\subsection{Interpretation: invariants and conservation laws}

\begin{figure}[h]
\begin{center}
\includegraphics[width=\columnwidth]{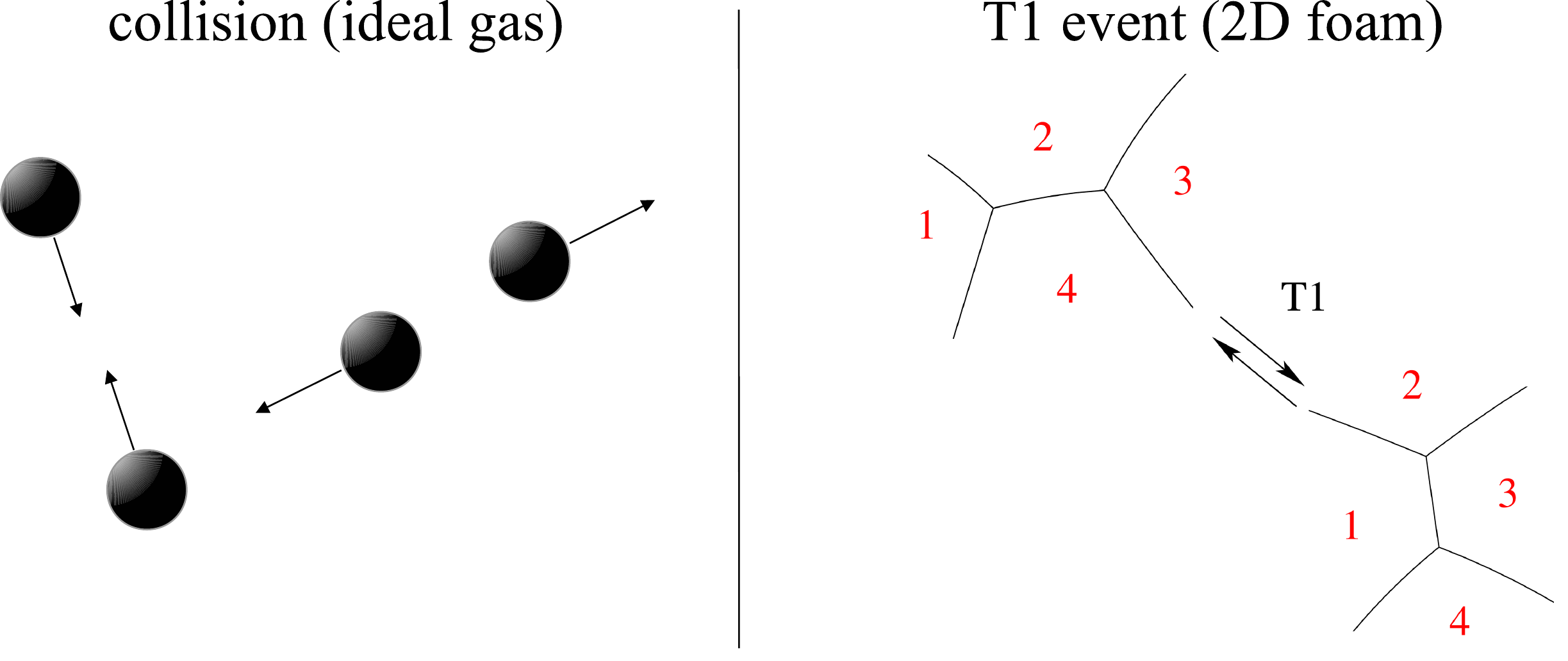}
\caption{analogy between a collision in an ideal gas and a T1 event in a 2D shuffled foam, playing the role of elementary ``thermalization'' process.}
\label{analogy}
\end{center}
\end{figure}

Self-consistency equations (\ref{constraints1}) and (\ref{constraints2}) can be interpreted as the conservation laws of the numbers of sides $n_i$ (or of the topological charge $q_t(n_i)$) and the cell curvatures $\kappa_\text{rb} (R_i,n_i)$, respectively. Since only the four bubbles directly involved in a T1 event have their $n_i$ and $\kappa_\text{rb} (R_i,n_i)$ which vary, these quantities are conserved locally.
A tight parallel can then be drawn between the dynamics of a shuffled foam and of an ideal gas, as illustrated in Fig. \ref{analogy}: T1 events in a shuffled foam play a role equivalent to the collisions between molecules in a gas. As for the momentum and energy of point particles, the total number of sides and bubble curvatures of the four bubbles involved in a $T1$ event are preserved, \textit{i.e.} they are the \textit{invariants} associated with the dynamics of the foam.
\revision{However, it must be emphasized that the local conservation of bubble curvature is valid only within the MFA, in which the bubble curvature depends primarily on $n_i$ and $R_i$, and not on the detail of its neighbourhood.} For a real foam, the space-filling constraints listed in Sect. \ref{section_constraints} generally imply a slight change of geometry of the surrounding bubbles as well. For instance, a T1 swap event in an initially regular hexagonal foam must induce a change of curvatures of sides in the neighbouring bubbles in order to satisfy the $120^\circ$ angle condition).

\subsection{Validity of Mean field approximation}
Mean field description requires that shuffling induces homogenization of the foam (no segregation of bubbles with same size), such that the number of sides of a bubble is decorrelated from its position within the foam. Experiments \cite{Vaz} and simulations \cite{Cox} support this hypothesis. 
%However, the homogeneization process of a shuffle foam will be the precise subject of a next study (ref Durand).

Mean field theories are known to work better when the number of neighbours increases, as their fluctuations average out. Therefore, one expects the model to give better results with bubbles having many sides (\textit{i.e.} with larger bubbles, statistically).
More generally, the mean field approximation consists in neglecting the fluctuations of bubble curvature. Thus, the approximation is expected to fail near the crystallization transition predicted by the model \cite{Durand3}. If this transition does exist, one expects to see an increase of T1 events close to it. Although this order-disorder transition is well identified in hard granular materials \cite{Stanley,Onuki1,Onuki2,Hilgenfeldt2}, it seems that it has never been studied numerically or experimentally in foams.

In summary, the mean field theory consists in \textit{i)} neglecting the fluctuations of neighbourhood: $\sum_{j \in \mathfrak{N}(i)}(\Pi_i-\Pi_j)$ depends essentially on $n_i$ and $R_i$, and barely on the specific foam configuration; and \textit{ii)} assimilating the mean bubble geometry with the geometry of a regular bubble. These assumptions are tested with Potts simulations (see Appendix B): Fig. \ref{cell-curvature-vs-n} shows the values of $R_i\widehat\kappa_i(\lbrace R_k \rbrace,n)$ in a tridisperse foam for different values of $n$. Tridispersity is a good compromise to have several bubble sizes while keeping a large number of bubbles of same size. 
To avoid superposition of hundreds of curves, $R_i\widehat\kappa_i(\lbrace R_k \rbrace,n)$ is calculated over all the bubbles having same size $R_i$.
For a given $n$, the values corresponding to the three different bubble sizes collapse very well, what confirms that this quantity depends on $n$ but not on $R_i$. Moreover, its variation with $n$ is in very good agreement with the regular bubble model [Eq. (\ref{regular_bubble_curvature})]. 
In the inset is plotted the dimensionless side curvature $R_i\widehat \kappa_i/n$ for the same data: they are aligned on a straight line passing through $0$ for $n=6$, in agreement with the regular bubble model [Eq. (\ref{regular_bubble_curvature})]. It can also be noticed that the \textit{relative} amplitude of the fluctuations decreases as $n$ increases, as expected.

\begin{figure}[h]
\begin{center}
\includegraphics[width=\columnwidth]{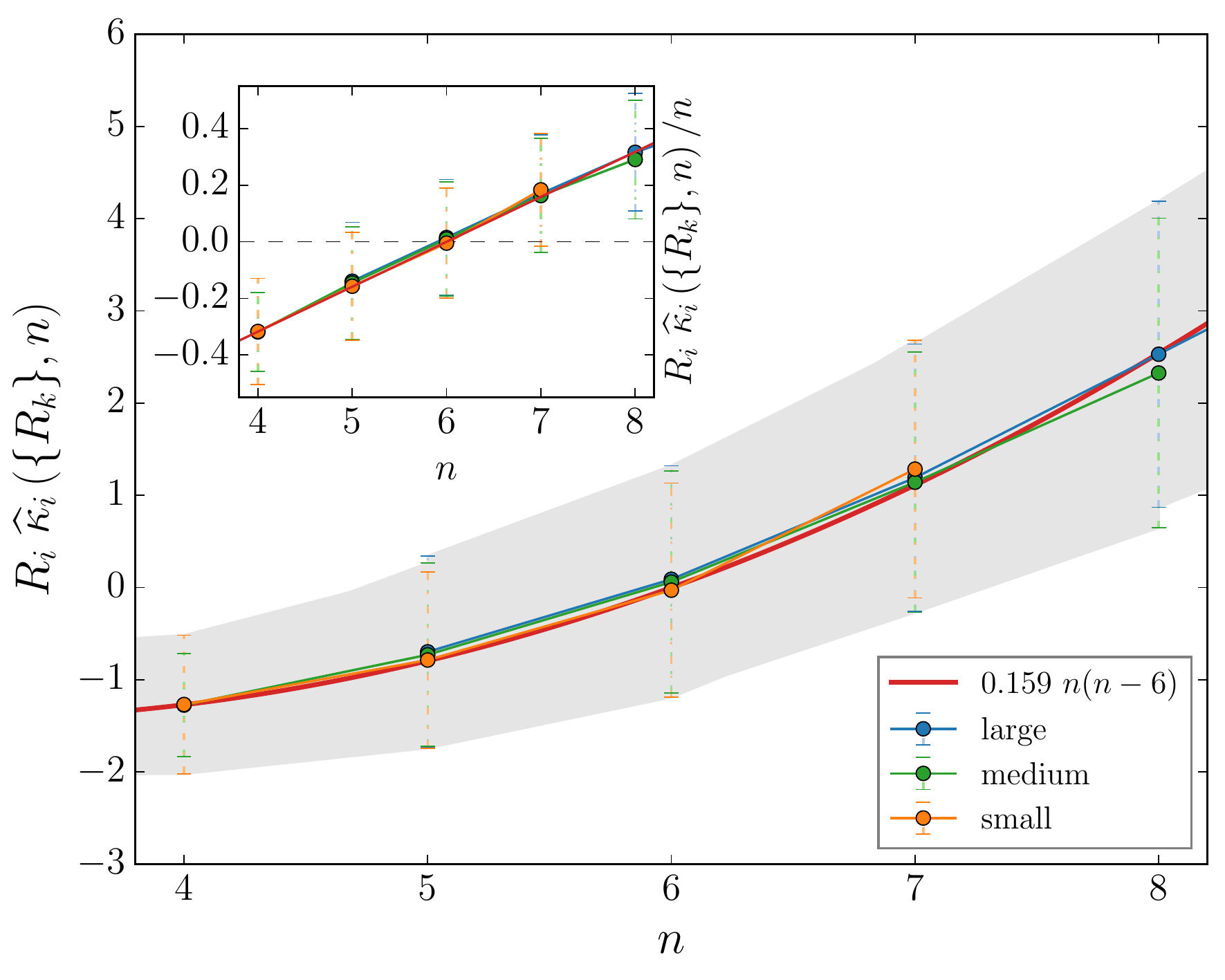}
\caption{Dimensionless mean bubble curvature $R \widehat\kappa$ \textit{vs.} $n$ in a tridisperse foam. Orange symbols: small bubbles (area$=200 ~ pixels^2$); green symbols: medium bubbles (area$=400 pixels^2$); blue symbols: large bubbles (area$=600 ~ pixels^2$). Proportion of small, medium and large bubbles are $0.319$, $0.335$ and $0.346$, respectively. Red curve: regular bubble model [Eq. (\ref{regular_bubble_curvature})]. Error bars and shaded area indicate the fluctuations (standard deviation) of  $R_i\sum_{j \in \mathfrak{N}(i)}(\Pi_{j}-\Pi_{i})/\gamma$ around its (partially) averaged value value $R_i\widehat\kappa_i(\lbrace R_k \rbrace,n)$. Inset: dimensionless mean side curvature $R \widehat\kappa(R,n)/n$ for the same data.}
\label{cell-curvature-vs-n}
\end{center}
\end{figure}

\begin{figure}[h]
\begin{center}
\includegraphics[width=\columnwidth]{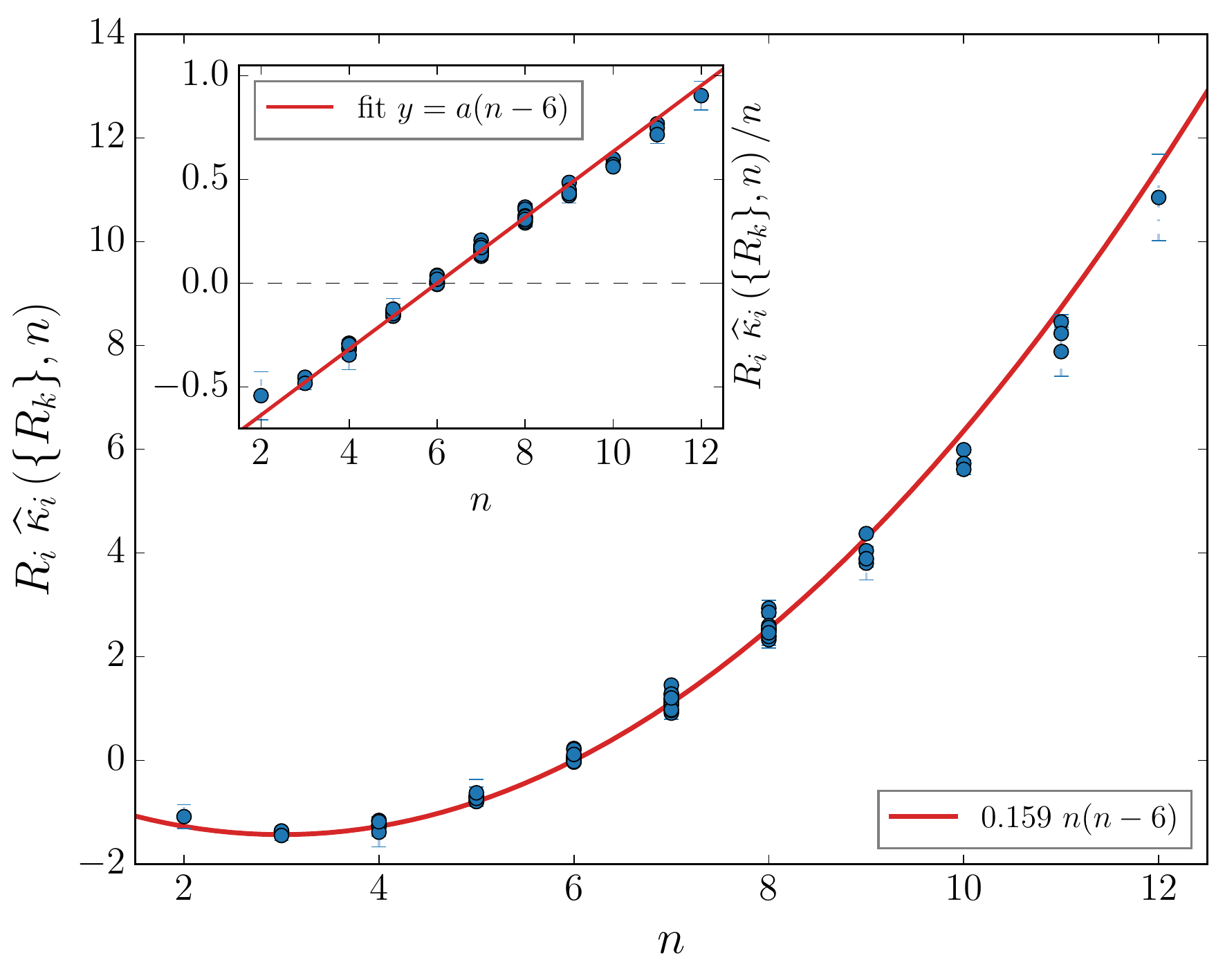}
\caption{Dimensionless mean bubble curvature $R\widehat\kappa$ \textit{vs.} $n$ for foams with various size dispersities: each point corresponds to an averaging over the bubbles with same $R$ and same $n$ within a given foam. Each foam contributes several points. Error bars represent the Standard Error of the Mean (SEM). Red curve: theoretical model [Eq. (\ref{regular_bubble_curvature})]. Inset: dimensionless mean side curvature $R_i\widehat\kappa_i/n$ \textit{vs.} $n$ for the same data. Red line: linear fit $y=a(n-6)$, with $a=0.157\pm0.004$.}
\label{cell-curvature-theo-vs-exp}
\end{center}
\end{figure}

Figure \ref{cell-curvature-theo-vs-exp} shows the same quantities compiled for foams with various size dispersities (bidisperse, tridisperse, and normal distributions of bubble areas). Once again, it shows a very good agreement with the regular bubble model [Eq. (\ref{regular_bubble_curvature})].
In particular, for a given value of $n$, the data corresponding to different foam polydispersity collapse very well, which confirms that the mean bubble curvature does not depend on the foam polydispersity, on the range of polydispersity tested ($\sigma_A/\langle A \rangle \leq 0.8$).
A linear fit (see inset) $a(n-6)$ to the mean side curvature yields $a=0.157 \pm 0.004$, which is very close from the value obtained in the regular bubble model ($a=\sqrt{\pi}/(3\epsilon)\simeq 0.159$).

It must be emphasized that although the MFA neglects the anti-correlations between the number of sides of neighbouring bubbles -- often modelized by the Aboav-Weaire law \cite{WeaireBook1,livre_mousse,Aboav,ODonovan,Oguey} -- it takes into account the correlations between the size and number of sides of every bubble.

\section{Ensemble equivalence} \label{section_stat_mech}

\subsection{Position of the problem}

In the mean field approximation, a microstate of the foam is specified by the position and number of sides of every bubble ($L\equiv\lbrace \mathbf{r}_i,n_i\rbrace$). 
Since the $\lbrace \mathbf{r}_i\rbrace$ and $\lbrace n_i\rbrace$ can be treated as independent variables (see Sect. \ref{section_dof}), the number of accessible states factorizes as $\Omega=\Omega^{(g)}\Omega^{(t)}$, where $\Omega^{(g)}$ is the number of configurations of the \textit{geometrical} variables $\lbrace \mathbf{r}_i\rbrace$, while $\Omega^{(t)}$ is the number of arrangements of the \textit{topological} variables $\lbrace n_i \rbrace$ that are compatible with the global constraints (\ref{constraints1}) and (\ref{constraints2}). $\Omega^{(t)}$ is then a function of the two \textit{state variables} $\K$ and $\N$. The other state variables are $\mathcal{N}_1,\dots,\mathcal{N}_p$, the number of bubbles with respective sizes $R_1,\dots,R_p$, (or equivalently, $\mathcal{N}$ and the bubble size distribution $P(R)$). For the sake of clarity, only $\K$ and $\N$ are made explicit hereafter.

In order to study the correlations between bubble size and side number, only $\Omega^{(t)}$ needs to be evaluated.
Let $\Omega^{(t)} \left(K_\mathscr{S},N_{\mathscr{S}}\vert \K,\N\right)$ be  the number of microstates of the foam for which the curvature and number of sides of a subset of bubbles $\mathscr{S}$ are $K_\mathscr{S}$ and $N_{\mathscr{S}}$, respectively.
The probability that $\mathscr{S}$ has $N_{\mathscr{S}}$ sides and curvature $K_\mathscr{S}$, given their total values $\N$ and $\K$ in the foam, is then:
\begin{equation}
P(K_\mathscr{S},N_{\mathscr{S}}\vert \K,\N)=\dfrac{\Omega^{(t)} \left(K_\mathscr{S},N_{\mathscr{S}}\vert \K,\N\right)}{\Omega^{(t)} \left( \K,\N\right)},
\label{condi_proba}
\end{equation}
However, the evaluation of $\Omega^{(t)}$
is not trivial (its derivation, above the crystallization, is reported on the Appendix C) and does not yields a simple expression for $P(K_\mathscr{S},N_{\mathscr{S}}\vert \K,\N)$. Instead, a ``grand-canonical description'' is proposed:
the ``small'' subset of bubbles $\mathscr{S}$ within the foam exchange sides and curvature with the rest of the foam through T1 events. The rest of the foam then constitutes a reservoir $\mathscr{R}$ of sides and curvature.
However, the micro- and grand-canonical descriptions are equivalent for large foams 
only if the system and the reservoir are \emph{weakly coupled}, that is, if the two following conditions are fulfilled:\\
\textit{i)} \textit{short-range correlations} between the bubbles %\{note1}
%subunits (bubbles)
to ensure the statistical independence of $\mathscr{S}$ and $\mathscr{R}$ \cite{Bertin2006,Chakraborty2010}. As discussed in Sect. \ref{correlations}, this is usually true in a real foam, except near the cristallization threshold where numerous T1s may occur. By definition, these correlations are neglected in the MFA. Therefore, $\Omega^{(t)}\left(N_{\mathscr{S}},K_\mathscr{S}\vert \N_{},\K\right)$ factorizes as
\begin{multline}
\Omega^{(t)}\left(K_\mathscr{S},N_{\mathscr{S}}\vert \K,\N\right)=\Omega^{(t)}_\mathscr{S}\left(K_\mathscr{S},N_{\mathscr{S}}\right) \times \\ \Omega^{(t)}_\mathscr{R}\left(K_\mathscr{R}=\K-K_\mathscr{S},N_{\mathscr{R}}=\N_{}-N_{\mathscr{S}}\right),
\label{factorization}
\end{multline}\\
where $\Omega^{(t)}_\mathscr{S}$ and $\Omega^{(t)}_\mathscr{R}$ refer to $\mathscr{S}$ and $\mathscr{R}$ when isolated, respectively.\\ 
\textit{ii)} \textit{extensivity} of the state variables that define a foam macrostate. Therefore, one must check the extensivity of $\N$ and $\K$.
This notion is often (mis)interpreted as a negligible contribution of the \textit{excess} (or \textit{coupling}) term when two systems (with at least one of macroscopic size) are put together \cite{long-range,Kuzemsky}. As an illustration, the energy of Hamiltonian systems is extensive when the interactions between the constituents are short-ranged.
\revision{Clearly this is not the case here for $\N$ and especially for $\K$, which has only an interface contribution: the coupling contribution to $\K$ between two bubble clusters that are mixed together is then of the same order of magnitude that their respective contributions, and thus cannot be neglected. However, this condition is unnecessary and one shows below that $\N$ and $\K$ are extensive quantities yet.}
Hence, the thermodynamic limit can be properly defined. The notion of reservoir is properly defined as well, \textit{i.e.} its effective temperature and chemical potential are not affected by the presence of the system $\mathscr{S}$ in contact.

\subsection{Extensivity and thermodynamic limit}\label{section_extensivity}

The scalings of $\N$ and $\K$ with $\mathcal{N}$ depend whether one considers a set of \textit{individual} bubbles (taken within a larger foam), or a cluster of bubbles. Clearly, in the first case, both variables scale linearly with $\mathcal{N}$.
%---------------------------
When bubbles are clustered, their is a nonlinear contribution of the outer boundary to the number of sides. From Euler's rule [Eq. (\ref{Euler})] and Plateau's law, the number of edges (Plateau borders) of a cluster is directly related to its number of bubbles: $N_{Pb}=3(\mathcal{N}+1-\chi)$. Since two adjacent bubbles share the same Plateau border, the number of sides is twice the number of Plateau borders, except for those at the outer boundary. Thus, the mean number of sides can be written $\N=6(\mathcal{N}+1-\chi)-c\sqrt{\mathcal{N}}$, where $c$ is a positive constant which depends on the size distribution within the cluster.

\begin{figure}[h]
\begin{center}
\includegraphics[width=0.4\columnwidth]{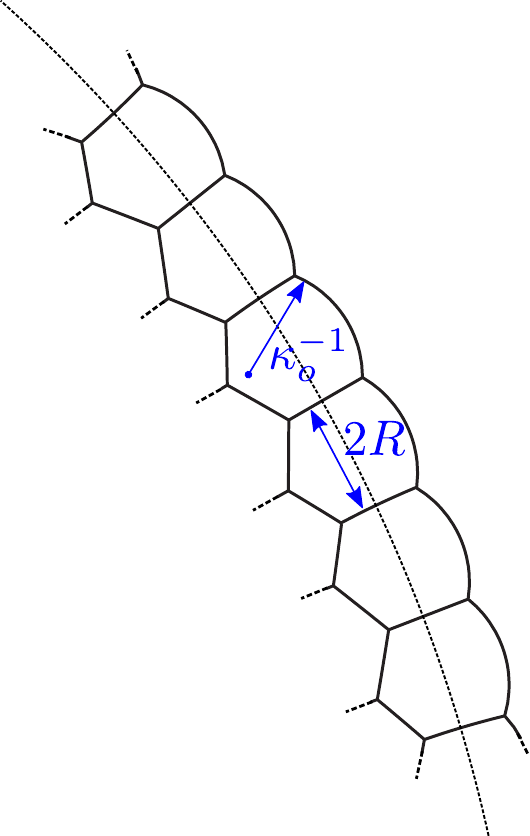}
\caption{Boundary of an idealized circular cluster made of identical bubbles.
}
\label{circular_cluster}
\end{center}
\end{figure}

While for $\N$ the contribution of the outer boundary becomes negligible in comparison with the bulk term as $\mathcal{N}$ increases, this is the one and only contribution for $\K$ (since side curvatures cancel in pairs in the bulk): $\K=-d\sqrt{\mathcal{N}}/\langle R \rangle$, where once again $d$ is a positive constant which depends on the size dispersion within the cluster.
For a foam enclosed in a container, $d=0$. For an unbounded foam, $c=d=0$. For a free cluster, an order of magnitude for $c$ and $d$ can be estimated considering a circular cluster (radius $R_c$) of identical bubbles with radius $R$, as shown on Fig. \ref{circular_cluster}. The perimeter of this cluster can be approximated as $2\pi R_c\simeq N_b(2 R)$, and its surface area as $\pi R_c^2\simeq\mathcal{N}(\pi R^2)$, where $N_b$ is the number of sides at the boundary. Therefore, $c=N_b/\sqrt{\mathcal{N}}\simeq\pi$. The radius of curvature $\kappa_0^{-1}$ of a side that belongs to the boundary tends to $2R$ when $\mathcal{N}\gg 1$. Therefore $d=N_b/\sqrt{\mathcal{N}} \kappa_0 R\simeq\pi/2$.

%--------------------------

Because of the boundary contribution, $\N$ and $\K$ are not additive quantities: when two shuffled clusters $\mathscr{S}_1$ and $\mathscr{S}_2$ -- with respective mean numbers of sides $\overline{N}_{1}^0$, $\overline{N}_{2}^0$ and respective mean curvature $\overline{K}_1^0$, $\overline{K}_2^0$ -- are mixed together, their total number of sides $\N_{1\cup 2}$ and total curvature $\K_{1\cup 2}$ are not equal to the sum of their values when they were isolated. As an illustration, consider two identical clusters, \textit{i.e.} made with the same number of bubbles ($\mathcal{N}_1=\mathcal{N}_2$) and same distribution of bubble size (hence $c_1=c_2$, $d_1=d_2$ and $\langle R \rangle_1=\langle R \rangle_2$), \revision{embedded in the 2D euclidean space ($\chi=2$).} Then 
\begin{equation}
\N_{1}^0=  \N_{2}^0= 6\left(\mathcal{N}_1 \revision{-1} \right)-c_1\sqrt{\mathcal{N}_1}
\end{equation}
and
\begin{equation}
\N_{1\cup 2}=  6(2\mathcal{N}_1\revision{-1})-c_1\sqrt{2\mathcal{N}_1}.
\end{equation}
The excess number of sides, defined as $\N_{12}=\N_{1\cup 2}-\N_{1}^0- \N_{2}^0$ growths sublinearly with $\mathcal{N}_1$: 
\begin{equation}
N_{12}=c_1(2-\sqrt{2})\sqrt{\mathcal{N}_1}\revision{+}6.
\end{equation}
Therefore, this contribution becomes negligible when $\mathcal{N}_1$ increases.
The mean curvature for each individual cluster is 
\begin{equation}
\K_2^0=\K_1^0=-d_1\sqrt{\mathcal{N}_1}/\langle R \rangle_1,
\end{equation}
and for the mixed clusters:
\begin{equation}
\K_{1\cup 2}=-d_1\sqrt{2\mathcal{N}_1}/\langle R \rangle_1.
\end{equation}
The excess curvature $\K_{12}=\K_{1\cup 2}-\K_1^0- \K_2^0$, is here of the same order of magnitude of the curvatures of the individual clusters:
\begin{equation}
\K_{12}=d_1(2-\sqrt{2})\sqrt{\mathcal{N}_1}/\langle R \rangle_1.
\end{equation}

Nevertheless, for separate bubbles as for clustered bubbles, $\N$ and $\K$ are \textit{extensive} quantities, meaning that $\N/\mathcal{N}$ and $\K/\mathcal{N}$ tend to finite limits as $\mathcal{N} \rightarrow \infty$ \cite{note-additivity-extensivity}. 
Moreover, fluctuations of $N$ and $K$ are always negligible for large systems: $\delta N/\N \rightarrow 0$, $\delta K/\K \rightarrow 0$ when $\mathcal{N} \rightarrow \infty$ (see Appendix A) These results together show that the thermodynamic limit exists for any large set of bubbles.
For clustered bubbles in particular, the limits of $N/\mathcal{N}$ ($ \simeq \N/\mathcal{N}$) and $K/\mathcal{N}$ ($\simeq \K/\mathcal{N}$) are universal, \textit{i.e.} they do not depend on $c$ and $d$, or on the specific bubble size distribution:
$\N/\mathcal{N} \rightarrow 6$ and $\K/\mathcal{N} \rightarrow 0$.
Eqs (\ref{constraints1}) and (\ref{constraints2}) can then be rewritten as: 
\begin{align}
\left< n \right> & = 6,\\
%\nonumber
\left<\frac{\sqrt{\pi} }{3} \frac{n(n-6)}{\epsilon R} \right> & = 0,
\end{align}
where $\left< \cdot \right>$ is the average defined in Sect. \ref{section_predictions}.

\begin{figure}[h]
\begin{center}
\includegraphics[width=\columnwidth]{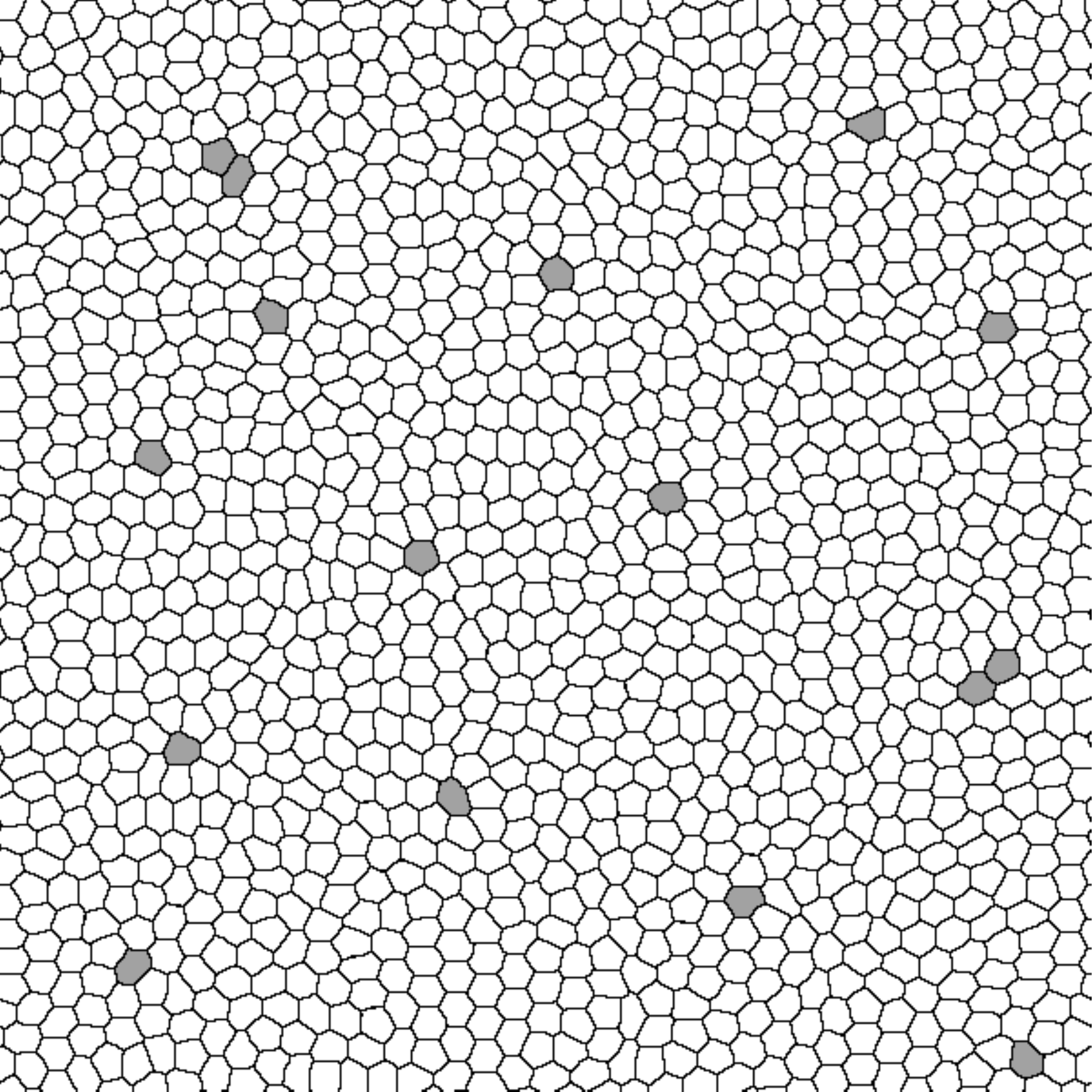}
\caption{Illustration of two foam samples $\mathscr{S}_1$ and $\mathscr{S}_2 \gg \mathscr{S}_1$ that are mixing together. At the initial stage, $\mathscr{S}_1$ (grey bubbles) was clustered.  Because of the agitation, bubbles do not remain clustered but scatter in $\mathscr{S}_2$ (white bubbles). $\mathscr{S}_2$ is still clustered but contains ``holes''.
}
\label{cluster_mixing}
\end{center}
\end{figure}

One can specify the conditions under which the larger system behaves as a reservoir of sides and curvatures for the smaller one: suppose that $\mathscr{S}_2$ is the larger system (\textit{i.e.} containing more degrees of freedom) and $\mathscr{S}_1$ is the smaller one.   
By definition, the values of effective temperature and chemical potential of a reservoir should not be affected by the presence of the smaller system, that is:

\begin{align}
%\begin{equation}
%\begin{split}
\deron{S_2}{K_2}_{\substack{K_2=\K_{1\cup 2}-\K_1 \\ N_{2} =\N_{1\cup 2}-\N_{1}}} & \simeq
%\deron{S_2}{K_2}_{\substack{\K_2=\K \\ \N_{2}=\N}}=
\deron{S_2}{K_2}_{\substack{K_2=\K_2^0 \\ N_{2}=\N_{2}^0}},
\label{reservoir1}
%\;\forall K_1,N_{1}, 
\\
\deron{S_2}{N_{2}}_{\substack{K_2=\K_{1\cup 2}-\K_1 \\ N_{2}=\N_{1\cup 2}-\N_{1}}} & \simeq
%\deron{S_2}{N_{2}}_{\substack{\K_2=\K \\ \N_{2}=\N}}=
\deron{S_2}{N_{2}}_{\substack{K_2=\K_2^0 \\ N_{2}=\N_{2}^0}}, \label{reservoir2}
%\deron{S_2}{N_{2}}_{\K,\N}=\deron{S_2}{N_2}_{\K-K_1,\N-N_{1}}\;\forall K_1,N_{1}
%\end{split}
%\end{equation}
\end{align}
\revision{where $S_2=\ln \Omega_2(K_2,N_2)$ is the entropy of $\mathscr{S}_2$.} As shown in Appendix C, the derivatives $\deronb{S_2}{K_2}(\K_2,\N_{2})$ and $\deronb{S_2}{N_{2}}(\K_2,\N_{2})$ involve the quantities $\N_{2}/\mathcal{N}_2$ and $\K_{2}/\mathcal{N}_2$. Therefore conditions (\ref{reservoir1}) and (\ref{reservoir2}) require $\N_{2}/\mathcal{N}_2\simeq \N_{2}^0/\mathcal{N}_2$ and $\K_{2}/\mathcal{N}_2\simeq \K_{2}^0/\mathcal{N}_2$ when $\mathcal{N}_2$ becomes large.
Under gentle shuffling, the whole system ($\mathscr{S}_{1\cup 2}$) remains clustered, so
\begin{equation}
\N_{1\cup 2}=6(\mathcal{N}_1+\mathcal{N}_2+1-\chi)-c\sqrt{\mathcal{N}_1+\mathcal{N}_2}
\end{equation}
and
\begin{equation}
\K_{1\cup 2}= -d\sqrt{\mathcal{N}_1+\mathcal{N}_2}.
\end{equation}
On the other hand, the number of configurations where the $\mathcal{N}_1$ bubbles are dispersed when the two systems are mixed is much larger than the number of configurations in which $\mathscr{S}_1$ remains clustered, as is illustrated in Fig. (\ref{cluster_mixing}). Thus, $\N_{1} \sim \mathcal{N}_1$ and $\K_{1} \sim \mathcal{N}_1$. Finally, $\mathscr{S}_2$ constitutes a cluster with $\mathcal{N}_1$ holes: $\N_{2} =\N_{1\cup 2}-\N_{1}$, and $\K_{2} =\K_{1\cup 2}-\K_{1} $.
Therefore, the condition under which $\mathscr{S}_2$ acts as a reservoir for $\mathscr{S}_1$ is: 
%$\mathscr{S}_2$ constitutes a reservoir when
\begin{equation} 
\sqrt{\mathcal{N}_2} \gg \mathcal{N}_1, 
\label{reservoir_condition}
\end{equation}
what is a more stringent condition than the one ($\mathcal{N}_2 \gg \mathcal{N}_1$) required for additive quantities (\textit{i.e.}: with no surface contributions) \cite{Books}.

\subsection{Grand-canonical description}\label{section:grand-canonical}
One can now legitimately use a grand-canonical ensemble to describe statistics of bubble shape. 
Without loss of generality, one considers that $\mathscr{S}$ is composed of a single bubble, with given size $R$ (alternatively, one could pick up a small subset of bubbles within the foam. That is the procedure that has been used in \cite{Durand1}). 
Thus,
\begin{align}
  \Omega^{(t)}_\mathscr{S}(K_\mathscr{S},N_{\mathscr{S}})=
  \begin{cases}
    1  & \text{if }  N_{\mathscr{S}}=n$ \text{and } $K_\mathscr{S}=\kappa_\text{rb}(R,n),\\
    0  & \text{otherwise. } 
  \end{cases}
\end{align}
In agreement with Eqs (\ref{condi_proba}) and (\ref{factorization}), the probability for the specified bubble to have $n$ sides is:
\begin{equation}
P(n\vert R)=\dfrac{\Omega^{(t)}_\mathscr{R} \left(K_\mathscr{R},N_\mathscr{R} \right)}{\Omega^{(t)}},
\label{proba}
\end{equation}
where $K_\mathscr{R}=\K-\kappa_\text{rb}(R,n)$ and $N_\mathscr{R}=\N-n$ are the respective curvature and number of sides of the reservoir $\mathscr{R}$, consisting of the $\mathcal{N}-1$ other bubbles.
The entropy of the reservoir associated with the topological degrees of freedom $\lbrace n_i \rbrace$ is defined as $S_\mathscr{R}=\ln \Omega^{(t)}_\mathscr{R}$. $S_\mathscr{R} \left(K_\mathscr{R},N_\mathscr{R} \right)$ can be expanded near $(\K,\N)$. As shown in Appendix C, successive terms in the expansion are smaller by a factor $1/\mathcal{N}$, as for additive variables \cite{Books}. Therefore:
\begin{multline}
S_\mathscr{R}\left(K_\mathscr{R}=\K-\kappa_\text{rb},N_\mathscr{R}=\N-n \right)= S_\mathscr{R}(\K,\N)
\\
 -\deronb{S_\mathscr{R}}{K_\mathscr{R}}
(\K,\N)\kappa_\text{rb}-\deronb{S_\mathscr{R}}{N_\mathscr{R}}
(\K,\N)n+\mathcal{O}\left(\mathcal{N}^{-2}\right),
\label{expansion_entropy}
\end{multline}
and thus, in the limit $\mathcal{N}\rightarrow \infty$:
\begin{equation}
\Omega^{(t)}_\mathscr{R} \left(K_\mathscr{R},N_\mathscr{R} \right)= \Omega^{(t)}_\mathscr{R} \left(\K,\N \right) e^{ -\beta \kappa(R,n) + \mu n },
\label{expansion_omega}
\end{equation}
with
\begin{equation}
\beta=\lim_{\mathcal{N}\rightarrow \infty} \deron{S_\mathscr{R}}{K_\mathscr{R}}
_{\substack{K_\mathscr{R}=\K \\ N_{\mathscr{R}}=\N}}
\label{beta_def}
\end{equation}
and
\begin{equation}
\mu=-\lim_{\mathcal{N}\rightarrow \infty}\left(\frac{\partial S_\mathscr{R}}{\partial N_\mathscr{R}}\right)
%\deron{S_\mathscr{R}}{N_\mathscr{R}}
_{\substack{K_\mathscr{R}=\K \\ N_{\mathscr{R}}=\N}}.
\label{mu_def}
\end{equation}
$\beta^{-1}$ and $\mu\beta^{-1}$ are the effective ``temperature'' and ``chemical potential'' of the reservoir, respectively. 

Eq. (\ref{proba}) eventually leads to the desired expression (\ref{historic}) for the conditional probability $P(n\vert R)$.
It can be noticed that one recovers the distribution intuited independently by 
Schliecker and Klapp \cite{Schliecker1,Schliecker2} and Sherrington and coworkers \cite{Sherrington1,Sherrington2}, but here with an explicit dependence on the bubble size $R$. This dependence 
is necessary to reflect the correlations between topological and geometrical properties.

\begin{figure}[h]
\begin{center}
\includegraphics[width=\columnwidth]{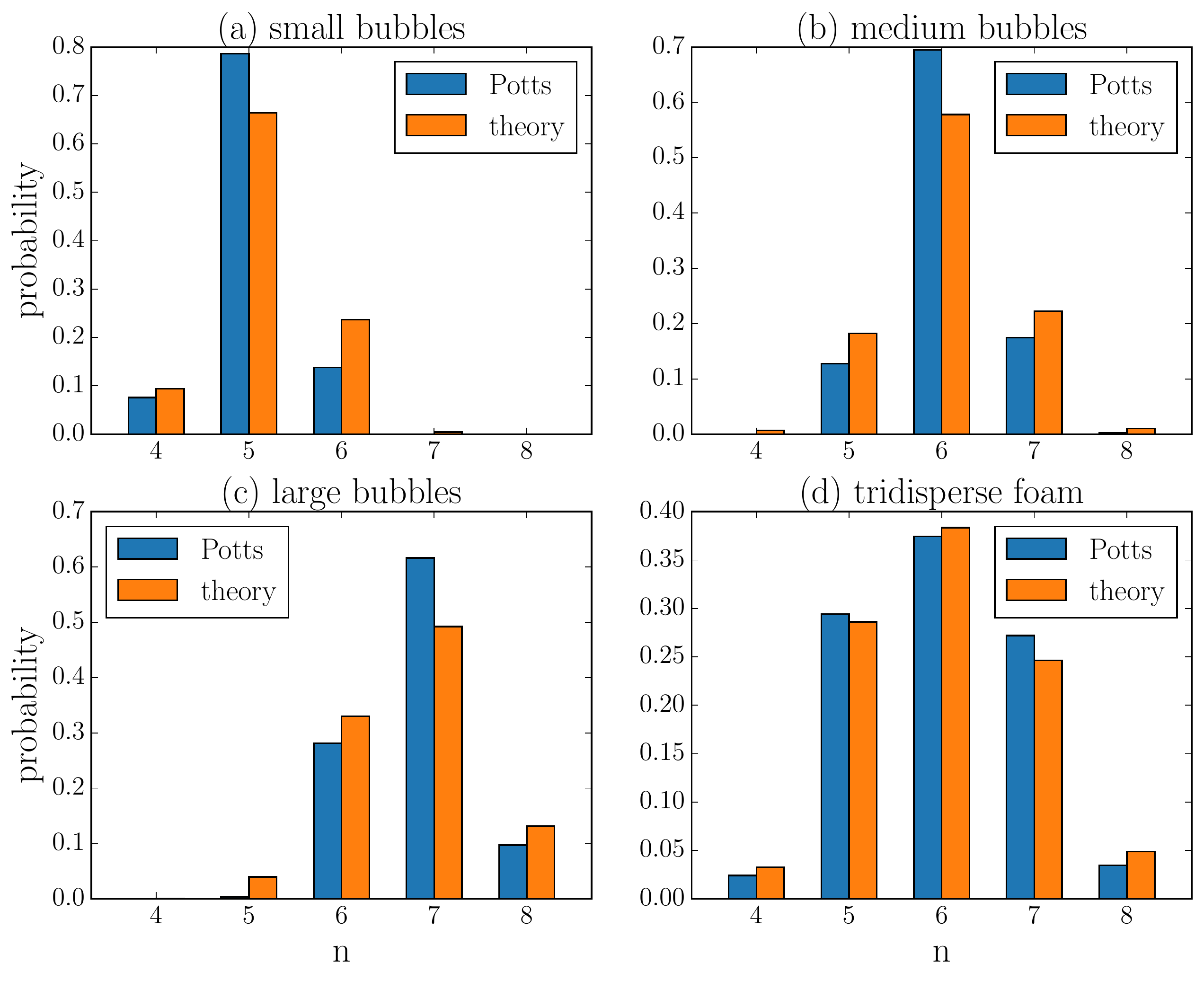}
\caption{\label{histogrammes}Theory \textit{vs} Simulations: conditional probability $P(n\vert R)$ for (a) small, (b) medium, and (c) large bubbles of a tridisperse foam (size ratios and number fractions are identical to those of Fig. \ref{cell-curvature-vs-n}; (d) distribution of number of sides $P(n)$ in the whole foam.}
\end{center}
\end{figure}

To illustrate the accuracy of the model, Figure \ref{histogrammes} compares the distributions predicted by the model with those obtained from Potts simulations, for the same tridisperse foam as in Fig. \ref{cell-curvature-vs-n}. The agreement is good for the conditional probabilities $P(n\vert R_\alpha)$ ($\alpha \in \lbrace 1,2,3 \rbrace$), and especially for the distribution of side number $P(n)=\sum_\alpha P(R_\alpha)P(n\vert R_\alpha)$. 

\section{Conclusions and perspectives}\label{section_outlook}

In summary, the different assumptions on which the statistical model is based have been systematically stated and argued in the present study, providing solids foundations to this theoretical framework.

This study also provides insights on possible improvement and extension of the model: foam coarsening could be taken into account to study the evolution of disorders on long time scales. The model can also be extended to biological tissues; in that case, shuffling can arise naturally from cell activity. Energy must be adapted, and cell division and apoptosis must be taken into account. 

Extension to 3D foams can be envisaged, but is not straightforward: unlike in 2D foams, the total number of sides (nor the total number of faces) of a foam that tiles the entire space is not imposed by the space-filling constraints in the limit $\mathcal{N} \rightarrow \infty$. Euler's rule combined with Plateau's laws just allow to relate the mean number of faces $\langle f \rangle$ to the mean number of sides $\langle n \rangle$ of a bubble \cite{WeaireBook1}:
\begin{equation}
\langle f \rangle=\dfrac{12}{6-\langle n \rangle}.
\end{equation}
$\langle f \rangle$ and $\langle n \rangle$ depend on the foam polydispersity \cite{Andy}. Moreover, a T1 event in a 3D foam is the transformation of a triangular face into a Plateau border. Consequently, not all the faces of a bubble have the opportunity to be involved in a T1, and the number of sides or faces are not invariants of a 3D shuffled foam. 

\begin{acknowledgement}
\textbf{Acknowledgments}\\
I wish to thank J. K\"afer for introducing me to the Cellular Potts simulations and F. Graner for insightful discussions. I am grateful to the CNRS for having benefited from a six months hosting program.
\end{acknowledgement}

%\appendix{
\section*{Appendix A: fluctuations}

\subsection*{Fluctuations of bubble areas, pressures, and curvatures}

Strictly, the pressure and area of a bubble fluctuate with configurations in a shuffled foam: only the number of gas molecules that it contains is conserved (in absence of foam coarsening). 
The variations of area and and pressure associated with a T1 event are imposed by the space filling constraints: consider for instance the transition from a regular, hexagonal foam to a new state obtained through a single T1 process. Gauss-Bonnet formula [Eq. \ref{Gauss-Bonnet}] require that the sides of the $5$ and $7$ cells are curved, which imply that the pressures in the bubbles are not identical anymore. A T1 event is not an isochoric or isobaric process. However, in simulations it is often easier to consider that bubble areas are preserved (incompressible gas). It is thus implicitly understood that the content of gas in the bubbles is not preserved in order to satisfy Young-Laplace and Plateau laws.

Pressure and area of any bubble $i$ are related through the equation of state of the gas, so their fluctuations have the same order of magnitude: typically $\delta \Pi_i/\overline{\Pi_i}=-g\delta A_i/\overline{A_i}$ with $g=\mathcal{O}(1)$ (\textit{e.g.} $g=1$ for an isothermal process, $g=3/2$ for an adiabatic process when bubbles contain a monoatomic ideal gas).
However, these fluctuations are generally very small: $\delta A_i/\overline{A_i} \sim \delta \Pi_i / \overline{\Pi_i} \ll 1$, and a bubble can be unambiguously identified by its area (which is then indistinguishable from its mean value $\overline{A_i}$). The criterion to discriminate the sizes of two bubbles $i$ and $j$ is that $\vert \overline{A_i}- \overline{A_j} \vert/(\overline{ A_i}+ \overline {A_j})$ is much larger than $\delta A_i/\overline{A_i}$ and $\delta A_j/\overline {A_j}$. 
 
It must be emphasized that although the pressure fluctuations are generally negligible ($\delta \Pi_i /\overline{\Pi_i} \ll 1$), the fluctuations or the pressure difference between two neighbouring bubbles $i$ and $j$
can be as large as $\vert ( \delta \Pi_i-\delta \Pi_j) / (\overline\Pi_i-\overline\Pi_j)\vert \sim \mathcal{O}(1)$. The variation is specially important when one of the bubbles, say bubble $i$, is directly involved in a T1 event. In the frame of our mean field approximation, 
\begin{equation}
\Pi_i-\Pi_j \simeq \gamma \kappa_\text{rb}(R_i,n_i)=\gamma \dfrac{\sqrt{\pi}}{3}\dfrac{\left(n_i-6 \right)}{\epsilon R_i}.
\end{equation}
Hence, the relative variation is $\simeq 1/(n_i-6)$, as $n_i$ increases or decreases by one.

\subsection*{Fluctuations of $K$ and $N$}
The fluctuations of the total curvature $K$ and total side number $N$ of a set of $\mathcal{N}$ bubbles depend on the specific boundary conditions: for an unbounded foam, both $K$ and $N$ are fixed and the fluctuations are trivially zero.

For a set of $\mathcal{N}$ individual bubbles taken in a larger foam, the application of central limit theorem on these $\mathcal{N}$ independent random variables yields: $\delta N/\N \sim \delta K/\K \sim \mathcal{N}^{-1/2}$. 

When the $\mathcal{N}$ bubbles are clustered, their side numbers an bubble curvatures are independent variables, because of the space-filling constraints. However, one can obtain the desired fluctuations with the following argument: let us ``cut'' the outer boundary into $b$ ($b \gg 1$) pieces. For a large cluster, the number of sides of each one of them are independent random variables, and then, using the central limit theorem again, the fluctuations of the total number of sides of the boundary $N_b$ satisfy $\delta N_b/\overline{N_b}\sim \overline{N_b}^{-1/2}$. Similarly, the fluctuations of the total curvature follow $\delta K/\K\sim \overline{N_b}^{-1/2}$. For a round cluster, $\overline{N_b}\sim \mathcal{N}^{1/2}$, and thus $\delta K/\K\sim \mathcal{N}^{-1/4}$. The total number of sides of the cluster, $N$, and those belonging to the outer boundary, $N_b$ are related by $N=6(\mathcal{N}+1-\chi)-N_b$. Therefore, for a large cluster ($\mathcal{N}\gg 1$), $\delta N/\N\sim \mathcal{N}^{-3/4}$. The decays of $\delta K/\K$ and $\delta N/\N$ are then different from those corresponding to a set of individual bubbles, and correct the scalings given in \cite{Durand1}. The scaling $\delta N/\N\sim \mathcal{N}^{-3/4}$ also holds for a foam enclosed in a container (while $\delta K/\K=0$).
In all cases, fluctuations become negligible as $\mathcal{N} \rightarrow \infty$.

\section*{Appendix B: Cellular Potts model}

The cellular Potts (or extended large-Q Potts) model has been applied to studies in several fields of physics or biology, among which: grain growth in metals, or coarsening in foams \cite{Glazier1990}, cell sorting \cite{Graner&Glazier}, or foam under shear \cite{Elias}.
The model discretizes the continuous cellular pattern onto a lattice, with an integer ``spin'' $\sigma_k$ defined at each lattice site $k$, chosen from $\lbrace 1,\dots,\mathcal{N} \rbrace$. Thus spin values merely act as labels for the $\mathcal{N}$ bubbles. Each bubble extends over many lattice sites: typically, each bubble in the simulations presented here covers approximately $300$ lattice sites. The Hamiltonian of the system reads:
\begin{equation}
\mathcal{H}=J\sum_{\substack{sites \\ \langle k,l \rangle}}\left(1-\delta_{\sigma_k,\sigma_l}\right)+\frac{B}{2}\sum_{\substack{bubbles \\ i}}\frac{\left(A_i-A_i^0\right)^2}{A_i^0}
\label{Hamiltonian}
\end{equation}
The first sum is carried over neighbouring sites $\langle k,l \rangle$ and represents the surface energy: each pair of neighbours having unmatching indices
determines a bubble wall and contributes to the bubble wall
surface energy. The surface tension $\gamma$ is then proportional to $J$, the coupling strength between neighbouring sites, with a prefactor which depends on the number of neighbours considered in the coupling. In the simulations presented here, the energy is evaluated with fourth nearest neighbour interactions (so each lattice pixel interacts with $20$ other pixels) to smooth the effect of lattice anisotropy \cite{Graner1}.

The second sum in Eq. (\ref{Hamiltonian}), carried over all the bubbles that constitute the foam, is the compressive energy of the bubbles. 
$B$ is the bulk modulus of the gas, and $A_i^0$ is the \textit{target area} of the bubble $i$, \textit{.i.e} the area that the enclosed gas would occupy if its pressure were equal to the surrounding pressure. It can be noticed that the expression of the elastic term differs slightly from the one commonly used in literature \cite{Graner1,Elias,Graner&Glazier}, because of the $A_i^0$ term in the denominator. This makes no difference for monodisperse foams (just a rescaling of the bulk modulus). Here, one must take the exact expression of compressive energy for bubbles with different sizes.
The system evolves using Monte Carlo dynamics. The algorithm slightly differs from the standard Metropolis algorithm: a site is chosen at random, but its value is reassigned only if it is at a bubble wall, and then \textit{only} to one of its unlike neighbours. This has two effects: one disregards unrealistic states in which bubble walls spontaneously emerge inside a bubble, and speed up computations. The algorithm used also forbids bubble fragmentation and satisfies the detailed balance equation \cite{Guesnet}.
The probability of accepting the trial reassignment is $P=1$ if the associated change of energy $\Delta \mathcal{H}$ is negative, and $P=\exp(-\Delta \mathcal{H}/\Theta)$ if $\Delta \mathcal{H} \geq 0$. $\Theta$ defines the \revision{temperature} of the Monte Carlo simulations. 

Parallel tempering technique \cite{Hukushima} has been used for simulating the foam at different values of temperature simultaneously. This technique allows to sample the foam dynamics efficiently at very low temperature. \revision{Values used in the present paper for the different parameters are $J=10$, $B=200$, and $\Theta=10$.}

\section*{Appendix C: microcanonical description}\label{Appendix2}
In this Appendix, the microcanonical entropy of a shuffled foam above the cristallization threshold is derived, \revision{within the MFA introduced in Sect. \ref{section_MFA}}. The expressions obtained from the grand-canonical description are recovered. Although a little bit trickier, the microcanonical description allows to establish the scaling of the successive derivatives of the entropy with $\mathcal{N}$, which is useful to give a quantitative definition of a reservoir [Eq. (\ref{reservoir_condition})] and to justify the expansion used in the grand-canonical description [Eq. (\ref{expansion_entropy})].
As in the Grand-canonical description, only $\Omega^{(t)}$ needs to be evaluated to study the correlations between topological and geometrical features. One treats bubbles of same size as discernable objects, the indiscernability factor being incorporated in $\Omega^{(g)}$. For simplicity, one assumes that $n_{\mathrm{min}}=3$. According to self-consistency equations (\ref{constraints1}) and (\ref{constraints2}), every accessible configuration must satisfy:
\begin{equation}
\sum_{i=1}^{\mathcal{N}}R_i^{1/2} x_i=\sqrt{\mathcal{N}\langle R \rangle}\delta, \qquad
%\nonumber
\sum_{i=1}^{\mathcal{N}} x_i^2=\mathcal{R}^2,
\label{constraints_cont} 
\end{equation}
with $x_i=\pi^{1/4}(n_i-3)/\sqrt{3 \epsilon R_i}$, $\delta=\pi^{1/4}(\N-3\mathcal{N})/\sqrt{3\epsilon \mathcal{N}\langle R \rangle}$ and $\mathcal{R}^2=3\sqrt{\pi}\mathcal{N} \langle R^{-1} \rangle/\epsilon+\K$. To get an analytic approximation of $\Omega^{(t)}$, one treats the $x_i$ (or $n_i$) as continuous variables.
\begin{figure}
\begin{center}
\includegraphics[width=0.8\columnwidth]{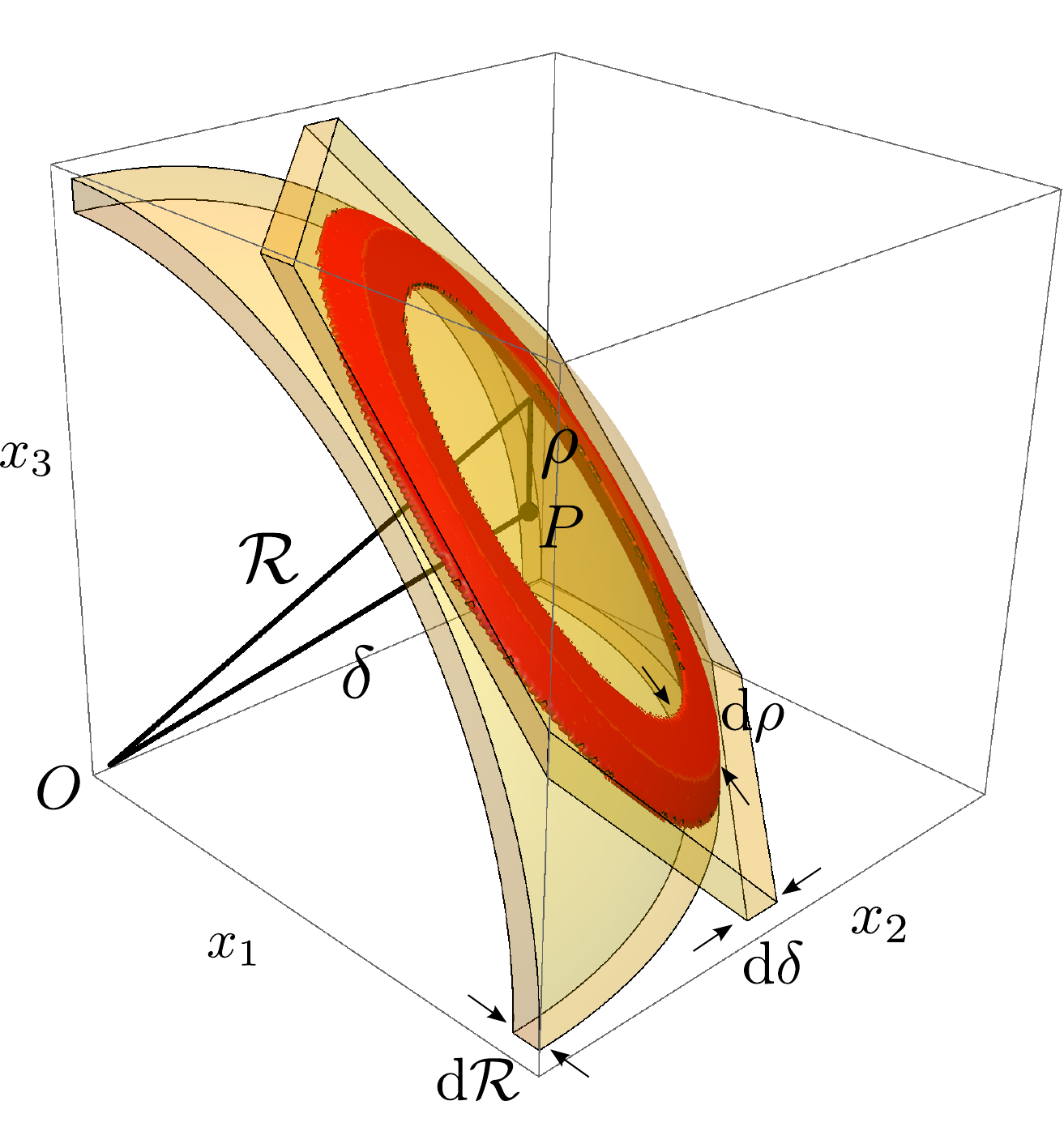}
\caption{Illustration in three dimensions ($\mathcal{N}=3$) of the volume occupied by the accessible states (in red) in the $x_i$-space. }
\end{center}
\label{3Dplot}
\end{figure}
Equalities (\ref{constraints_cont}) are respectively the equations of a $\mathcal{N}-$hyperplane lying at a distance $\delta$ from the origin $O$, and of a $\mathcal{N}-$hypersphere of radius $\mathcal{R}$ centered on $O$.
Since $\K \sim -\sqrt \mathcal{N}$ and $\N\geq 3\mathcal{N} \gg 1$, one has $0\leq\delta\leq\mathcal{R}$.
The intersection of these two surfaces is the $(\mathcal{N}-1)-$ hypersphere with radius $\rho=\sqrt{\mathcal{R}^2-\delta^2}$ and centered on $P$, the projection of $O$ on the hyperplane (see Fig. \ref{3Dplot}). Its surface is $S_{\mathcal{N}-1}(\rho)=2\pi^{(\mathcal{N}-1)/2}\rho^{\mathcal{N}-2}/\Gamma\left[(\mathcal{N}-1)/2\right]$, where $\Gamma$ is the Euler's Gamma function. 
$S_{\mathcal{N}-1}(\rho)\dd \rho \dd \delta$ is the hypervolume of the region defined by the intersections of the hypershell of thickness $\dd \rho=\mathcal{R}\dd\mathcal{R}/\rho=\dd\K/(2\rho)$ with the two hyperplanes lying at a distance $\delta$ and $\delta+ \dd \delta$ from $O$ respectively, with $\dd \delta=\pi^{1/4}\dd \N/\sqrt{3\epsilon \mathcal{N}\langle R \rangle}$.
Within the continuous approximation, the number of states $\Omega^{(t)}(\K,\N;\dd \K,\dd \N)$ whose total curvature and number of sides lie between $\left[ \K,\K+\dd \K\right[$ and  $\left[ \N,\N+\dd \N \right[$, respectively, is simply the ratio of this volume to the hypervolume occupied by one microstate, $v=\prod_{i=1}^\mathcal{N} \delta x_i$, with $\delta x_i=\pi^{1/4}/\sqrt{3\epsilon R_i}$. It yields:
\begin{multline}
%\begin{split}
\Omega^{(t)}(\K,\N;\dd \K,\dd \N) =   %\nonumber
\\
%\left(\prod_{i=1}^\mathcal{N} R_i^{1/2}\right)
\e^{\mathcal{N}\langle \ln R \rangle/2}
\frac{2}{\sqrt{\pi}}\frac{(3\epsilon\sqrt{\pi})^{\mathcal{N}/2}\rho^\mathcal{N}}{\Gamma\left(\frac{\mathcal{N}-1}{2}\right)}\frac{\dd\rho}{\rho}\frac{\dd\delta}{\rho}. 
%\end{split}
\end{multline}
$\dd\rho/\rho$ and $\dd\delta/\rho$ are slowly varying functions of $\mathcal{N}$. Using Stirling approximation for $\Gamma$ when $\mathcal{N}\gg 1$, the expression of the entropy of the foam $S(\K,\N)=\ln \Omega^{(t)}$ simplifies to: 
\begin{multline}
S(\K,\N)\simeq\frac{\mathcal{N}}{2}\left\lbrace \ln\left[ 2\e\sqrt{\pi} \left( 9\langle R^{-1} \rangle \phantom{\left(\frac{\N}{\mathcal{N}}-3\right)^2} \right. \right. \right.  \\ 
\left. \left. \left. -\left(\frac{\N}{\mathcal{N}}-3\right)^2 \langle R \rangle^{-1}\right)+6\e\epsilon\frac{\K}{\mathcal{N}} \right]+\left\langle \ln R \right\rangle \right\rbrace ,
\label{entropy}
\end{multline}
where $\e$ is the Euler's number.
In the limit $\mathcal{N} \rightarrow \infty$, $\K/\mathcal{N}$ and $\N/\mathcal{N}$ tend to constant values (see Sect. \ref{section_extensivity}), and the entropy becomes extensive. In particular, for a reservoir (that is, a large bubble cluster), $\K/\mathcal{N} \rightarrow 0$ and $\N/\mathcal{N} \rightarrow 6$, and the entropy reads: 
%in the limit $\mathcal{N} \rightarrow \infty$:
\begin{equation}
S_\mathscr{R}=\frac{\mathcal{N}}{2}\left\lbrace \ln\left[ 18\e\sqrt{\pi} \left( \langle R^{-1} \rangle- \langle R \rangle^{-1}\right)\right]+\langle \ln R \rangle \right\rbrace.
\label{extensive_entropy}
\end{equation}

One can now justify the limitation up to first order used in the grand-canonical description [Eq. (\ref{expansion_entropy})] for the expansion of the entropy of the reservoir $S_\mathscr{R}\left(K_\mathscr{R},N_\mathscr{R}\right)=\ln \Omega_\mathscr{R} \left(K_\mathscr{R},N_\mathscr{R} \right)$ near $(\K,\N)$:
using Eq. (\ref{entropy}), it is easy to see that, as $\mathcal{N}\rightarrow \infty$,
\begin{equation}
\left( \dfrac{\partial^{p+q}S_\mathscr{R}}{\partial K^p_\mathscr{R} \partial N^q_\mathscr{R}}\right)_{\substack{K_\mathscr{R}=\K \\ N_\mathscr{R}=\N}} \sim \mathcal{N}^{1-(p+q)},
\end{equation}
for any integers $p,q \geq 0$. Therefore, successive terms in the expansion are smaller by a factor $1/\mathcal{N}$, as it is for additive quantities. Note that such a scaling is not mandatory when the entropy depends on non-additive quantities as it is the case here.
In particular, the first derivatives are: 
\begin{align}
\deron{S_\mathscr{R}}{K_\mathscr{R}}_{\substack{K_\mathscr{R}=\K \\ N_\mathscr{R}=\N}} & = \frac{3\e\epsilon}{Q},
\\
\deron{S_\mathscr{R}}{N_\mathscr{R}}_{\substack{K_\mathscr{R}=\K \\ N_\mathscr{R}=\N}} & = -\left(\frac{\N}{\mathcal{N}}-3\right)\frac{2\e\sqrt{\pi}}{\langle R \rangle Q},
\end{align}
with 
\begin{equation}
Q=2\e\sqrt{\pi} \left( 9\langle R^{-1} \rangle-\left(\frac{\N}{\mathcal{N}}-3\right)^2 \langle R \rangle^{-1}\right)+6\e\epsilon\frac{\K}{\mathcal{N}}.
\end{equation}
The two derivatives become intensive quantities in the limit $\mathcal{N}\rightarrow \infty$. In agreement with Eqs (\ref{beta_def}) and (\ref{mu_def}), the asymptotic values of the two derivatives define the temperature and chemical potential of the reservoir. 
The reader can easily check that their expressions are identical to those obtained from the grand-canonical description [Eqs (\ref{beta}) and (\ref{mu})] above the crystallization threshold. This is not a surprise, as both derivations use the same approximation, in which the $\lbrace n_i \rbrace$ are treated as continuous variables \cite{Durand2,Durand3}.
%This expansion requires that the higher terms are negligible.

The grand partition function $\Xi$ is related to $S$ by:
\begin{equation}
\ln \Xi=S-\beta\K+\mu \N.
\end{equation}
Hence, maximization of $S$ under fixed values of $\K$ and $\N$ is formally equivalent to the minimization of the \textit{Grand Potential} $\mathcal{J}=-\beta^{-1} \ln \Xi$ (under fixed values of $\beta$ and $\mu$).

Note that the continuous approximation requires that, for every $i$, $\delta x_i$ is much smaller than $h_i$ and $\mathcal{R}$, \textit{i.e.} $\mathcal{N}\gg1$ and $\mathcal{N}\langle R^{-1}\rangle\gg R_i^{-1}$. These conditions are ensured whenever there is a large number of bubbles with same size. 
Moreover, the number of accessible states $\Omega^{(t)}$ must be large, which requires $\e^{\langle \ln R \rangle}18\e \pi \left( \langle R^{-1} \rangle- \langle R \rangle^{-1}\right)>1$ for large $\mathcal{N}$. This is a lower bound to the range of polydispersity on which the continuous approximation is valid. Indeed, at lower dispersity, the model predicts a crystallization transition \cite{Durand3}.
There is also an upper limit to the range of validity of the continuous approximation: most of the intersection between the two hypersurfaces must lie in the space domain $\lbrace x_i \geq 0, i=1,\dots,\mathcal{N}\rbrace$ to satisfy the constraints $n_i\geq 3$. The proportion of the intersection lying in this space domain is not easy to calculate. Numerically, it has been established that the condition is satisfied whenever $\langle R \rangle \langle R^{-1} \rangle- 1\lesssim 0.35$ and $ (\sigma_n / \langle n \rangle)^2 \lesssim 0.4$ \cite{Durand3}.

%\section{bibliography}


\begin{thebibliography}{70}

	\bibitem{Durand1} M. Durand, \textit{EPL (Europhysics Letters)} \textbf{90}, 60002 (2010).

	\bibitem{Durand2} M. Durand, J. K\"afer, C. Quilliet, S. J. Cox, S. Ataei Talebi, and F. Graner, \textit{Physical Review Letters}, \textbf{107}, 168304 (2011).

	\bibitem{Durand3} M. Durand, A. Kraynik, F. Van Swol, J. K\"afer, C. Quilliet, S. J. Cox, S. Ataei Talebi, and F. Graner, \textit{Physical Review E} 062309, \textbf{89} (2014).

	\bibitem{Edwards-and-co} S. F. Edwards and R. B. S. Oakeshott, \textit{Physica A}, \textbf{157}, 1080 (1989).
%

	\bibitem{Blumenfeld} R. Blumenfeld and S. F. Edwards,  {\it Phys. Rev. Lett.} {\bf 90}, 114303 (2003).

	\bibitem{WeaireBook1} D. L. Weaire and S. Hutzler, \textit{The Physics of Foams}, Oxford University Press (2000).
%
	\bibitem{livre_mousse} I. Cantat, S. Cohen-Addad, F. Elias, F. Graner, R. H\"ohler, O. Pitois, F. Rouyer, A. Saint-Jalmes  (ed. S.J. Cox).  \textit{Foams: structure
and dynamics}, Oxford University Press, Oxford (2013).

\bibitem{Schliecker1} G. Schliecker, \textit{Adv. Phys.} \textbf{51},1319 (2002).

\bibitem{Schliecker2} G. Schliecker and S. Klapp, \textit{Europhys. Lett.} \textbf{48}, 122-128 (1999).

\bibitem{Rivier1} N. Rivier, A. Lissowski, \textit{J. Phys. A} \textbf{15}, L143 (1982).

\bibitem{Rivier2} N. Rivier, \textit{Phil. Mag. B} \textbf{52}, 795-819 (1985).

\bibitem{Fortes3} M. A. Fortes, P. I. C. Texeira, \textit{J. Phys. A} \textbf{36}, 5161 (2003).

\bibitem{Iglesias} J. R. Iglesias, R. M. C. de Almeida, \textit{Phys. Rev. A} \textbf{43}, 2763 (1991).

\bibitem{sire} C. Sire and M. Seul, {\it J. Phys. I France} {\bf 5} 97 (1995).

\bibitem{Quilliet}  C. Quilliet, S. Ataei Talebi, D. Rabaud, %%{\it et al.},
J. K\"afer, S. J. Cox. and F. Graner, 
{\it Phil. Mag. Lett.} {\bf 88} 651 (2008).

\bibitem{PNAS} J. K\"{a}fer, T. Hayashi, A. F. M. Mar\'{e}e, R. W. Carthew, and F. Graner, \textit{PNAS} \textbf{104}, 18549-18554 (2007); S. Hilgenfeldt, S. Eriksen, and R. W. Carthew, \textit{PNAS} \textbf{105}, 907-911 (2008).

\bibitem{Nagel&Liu} A. J. Liu and S. R. Nagel, \textit{Nature} \textbf{396}, N6706, 21 (1998).
%


\bibitem{Barrat1} A. Nicolas, K. Martens, and J.-L. Barrat, \textit{EPL} \textbf{107},  44003 (2014).

\bibitem{Barrat2} E. Agoritsas, E. Bertin, K. Martens, J.-L. Barrat, submitted to \textit{Eur. Phys. J. E.}

\bibitem{Euler} D. S. Richeson, \textit{Euler's Gem: The Polyhedron Formula and the Birth of Topology}, Princeton University Press, 2008.

\bibitem{Durand&Stone} M. Durand and H. A. Stone, \textit{Phys. Rev. Lett.} \textbf{97}, 226101 (2006).

\bibitem{Bowles}To jump from one metastable state to another one which occupies the same volume, a packing of hard grains must temporarily increase its volume: see R. K. Bowles, S. S Ashwin, \textit{Physical Review E}, \textbf{83}, 031302 (2011).

\bibitem{note_uniform_strain}\revision{However, simple shear with displacements imposed on all the frame boundaries, as in \cite{Quilliet}, induces a very homogeneous deformation.}

\bibitem{Dennin} Y. Wang, K. Krishan, and M. Dennin, \textit{Phys. Rev. Lett.} \textbf{98}, 220602 (2007); \textit{id.}, \textit{Phys. Rev. E} \textbf{73}, 031401 (2006).

	\bibitem{shear-banding} G. Katgert, M. E. Mobius, and M. van Hecke PRL 101, 058301 (2008).

	\bibitem{Books} F. Reif, \textit{Fundamentals of Statistical Physics and Thermal Physics}, McGraw-Hill (1965).
L.D Landau and E. M. Lifshitz, \textit{Statistical Physics}, Pergamon Press (1969).

	\bibitem{Jiang} Y. Jiang, P. J. Swart, A. Saxena, M. Asipauskas, and J. A. Glazier, \textit{Phys. Rev. E} \textbf{59}, 5819-5832 (1999).

	\bibitem{Graner1} Graner, F., Jiang, Y., Janiaud, E., and Flament, C., \textit{Phys. Rev. E} \textbf{63}, 011402 (2001).

%	\bibitem{Jendrol} S. Jendrol' and E. Jucovi\v{c}, \textit{Proc. London Math. Soc.} (3) \textbf{25},
385–398 (1972).

%	\bibitem{Kussner} I Izmestiev \textit{et al.}, \textit{Geometriae Dedicata}\textbf{166}, 15-29 (2013).

	\bibitem{Nelson} \textit{Statistical Mechanics of Membranes and Surfaces}, eds. D. Nelson, T. Piran, and S. Weinberg (World Scientific, Teaneck, NJ, 1989).

	\bibitem{Fortes1} M.A. Fortes, P.I.C. Teixeira,\textit{ Eur. Phys. J. E} \textbf{6}, 255-258 (2001).
%

	\bibitem{Weaire2} D. Weaire, S.J. Cox, F. Graner, \textit{Eur. Phys. J. E} \textbf{7}, 123-127 (2002).

	\bibitem{Vaz3} M.F. Vaz and M.A. Fortes, \textit{Eur. Phys. J. E} \textbf{11}, 95-97 (2003).

	\bibitem{Moukarzel} C. Moukarzel, \textit{Phys. Rev. E} \textbf{55} 6866 (1997).

	\bibitem{note_fluctuations} Moreover, bubble areas fluctuate (see Appendix A), yielding another source of uncertainty for their geometrical centers.

	\bibitem{Cox2} S. J. Cox, F. Graner and M. F. Vaz, \textit{Soft Matter} \textbf{4}, 1871-1878 (2008).

	\bibitem{Elias} F. Elias, C. Flament, J. A. Glazier, F. Graner, Y. Jiang, \textit{Philos. Mag. B} \textbf{79} 729-751 (1999).

	\bibitem{Durian1991} D. J. Durian, D. J. Pine, D. A. Weitz, \textit{Science} \textbf{252}, 686 (1991).
%

	\bibitem{Duri2009} A. Duri, D. A. Sessoms, V. Trappe, L. Cipelletti, \textit{Phys. Rev. Lett.}, \textbf{102}, 085702 (2009).
%

%	\bibitem{Iglesias} J. R. Iglesias, R. M. C. de Almeida, \textit{Phys. Rev. A} \textbf{43}, 2763 (1991).


	\bibitem{Aboav} D. A. Aboav, \textit{Metallography} \textbf{3}, 383 (1970).

	\bibitem{ODonovan} C.B. O'Donovan, E.I. Corwin, M.E. M\"obius, \textit{Philosophical Magazine} \textbf{93}, 31-33, 4030-4056 (2013).

	\bibitem{Oguey} M.F. Miri, C. Oguey, \textit{Colloids and Surface A}, \textbf{309}, 107-111 (2007).

	\bibitem{Vaz} M. F. Vaz, S.J. Cox, P.I.C. Teixeira, \textit{Phil. Mag.} \textbf{91}, 4345-4356 (2011).

	\bibitem{Cox} S.J. Cox, \textit{J. non-Newt. Fl. Mech.} \textbf{137}, 39-45 (2006).

	\bibitem{Stanley} M.R. Sadr-Lahijany, P. Ray, H.E. Stanley, \textit{Phys. Rev. Lett.} \textbf{79} (1997) p.3206.

	\bibitem{Onuki1} T. Hamanaka and A. Onuki, \textit{Phys. Rev. E} \textbf{74} (2006) p.011506.

	\bibitem{Onuki2} T. Hamanaka and A. Onuki, \textit{Phys. Rev. E} \textbf{75} (2007) p.041503.

	\bibitem{Hilgenfeldt2} S. Hilgenfeldt, \textit{Philosophical Magazine} \textbf{93}, 4018--29 (2013).

%

	\bibitem{Bertin2006} E. Bertin, O. Dauchot, M. Droz, \textit{Phys. Rev. Lett.}, \textbf{96}, 120601 (2006).
%

	\bibitem{Chakraborty2010} B. Chakraborty, \textit{Soft Matter} \textbf{6}, 2884-2893 (2010).
%

	\bibitem{long-range} T. Dauxois, S. Ruffo, E. Arimondo and M. Wilkens (Eds.), \textit{Dynamics and Thermodynamics of Systems with Long-Range Interactions}, Lecture Notes in Physics \textbf{602}, Springer-Verlag (2002).

	\bibitem{Kuzemsky} {A. L. Kuzemsky, \textit{International Journal of Modern Physics B} \textbf{28}, p.1430004 (2014).}
%

	\bibitem{note-additivity-extensivity} Extensivity does not imply additivity. The converse is true for homogeneous systems only.


	\bibitem{Sherrington1} T. Aste and D. Sherrington, \textit{J. Phys. A: Math. Gen.} \textbf{32}, 7049-7056 (1999).

	\bibitem{Sherrington2} L. Davison and D. Sherrington, \textit{J. Phys. A: Math. Gen.} \textbf{33}, 8615-8625 (2000).

	\bibitem{Andy} A. M. Kraynik, D. A. Reinelt, F. van Swol, \textit{Phys. Rev. Lett.} \textbf{93} 208301 (2004).


	\bibitem{Glazier1990} J. A. Glazier, M. P. Anderson, G. S. Grest, \textit{Philos. Mag. B} \textbf{62}, 615-645 (1990).

	\bibitem{Graner&Glazier} F. Graner and J. A. Glazier, \textit{Phys. Rev. Lett.} \textbf{69}, 2013-2016 (1992); J. A. Glazier and F. Graner, \textit{Phys. Rev. E} \textbf{47}, 2128-2154 (1993).

	\bibitem{Guesnet} E. Guesnet and M. Durand, \textit{in preparation}.

	\bibitem{Hukushima} K. Hukushima and K. Nemoto, \textit{J. Phys. Soc. Jap.} \textbf{65}, 1604-1608 (1996).


\end{thebibliography}
\end{document}